\documentclass[a4paper,fleqn]{cas-sc}

\usepackage[numbers]{natbib}
\usepackage{timet}
\usepackage{graphicx}
\usepackage{multirow}
\usepackage{multicol}
\usepackage{lipsum}
\usepackage{timet}
\usepackage{epsfig}
\usepackage{amsmath}
\usepackage{amsfonts}
\usepackage{amssymb}
\usepackage{color}
\usepackage{booktabs}
\usepackage[overload]{empheq}

\usepackage{float}
\usepackage{subcaption}
\usepackage{graphics}
\usepackage{xcolor,graphicx}
\usepackage{csquotes}
\usepackage{mathtools}
\usepackage{optidef}
\usepackage{hyperref}

\usepackage{algorithm}
\usepackage{algpseudocode}
\usepackage{rotating}
\usepackage{multirow}
\usepackage{adjustbox}
\usepackage{lscape}

\definecolor{linkcol}{RGB}{0,128,172}
\hypersetup{
  colorlinks = true,
  linkcolor  = linkcol,   
  citecolor  = linkcol,   
  urlcolor   = linkcol,   
  filecolor  = linkcol,
}

\usepackage{mathtools}
\usepackage[normalem]{ulem}
\usepackage{cancel}

\usepackage[nameinlink]{cleveref}
\crefname{figure}{Fig.}{Figs.}          \Crefname{figure}{Figure}{Figures}
\crefname{table}{Table}{Tables}         \Crefname{table}{Table}{Tables}
\crefname{section}{Section}{Sections}   \Crefname{section}{Section}{Sections}
\crefname{algorithm}{Algorithm}{Algorithms}
\Crefname{algorithm}{Algorithm}{Algorithms}
\crefname{equation}{Eq.}{Eqs.}
\Crefname{equation}{Equation}{Equations}

\crefname{appendix}{Appendix}{Appendices}
\Crefname{appendix}{Appendix}{Appendices}

\newcommand{\bx}{\mathbf{x}}
\DeclareMathOperator*{\argmin}{arg\,min}  
\DeclareMathOperator*{\minimize}{minimize}
\DeclareMathOperator*{\maximize}{maximize}
\def\tsc#1{\csdef{#1}{\textsc{\lowercase{#1}}\xspace}}
\tsc{WGM}
\tsc{QE}

\begin{document}
\let\WriteBookmarks\relax
\def\floatpagepagefraction{1}
\def\textpagefraction{.001}

\renewcommand{\topfraction}{0.9}
\renewcommand{\bottomfraction}{0.8}
\renewcommand{\textfraction}{0.07}
\renewcommand{\floatpagefraction}{0.7}
\setcounter{topnumber}{3}
\setcounter{totalnumber}{5}

\newcommand{\sgrev}[1]{{\color{blue} #1}}
\newcommand{\newtxt}[1]{{\color{magenta} #1}}

\newcommand{\delete}[1]{\textcolor{red}{\sout{#1}}}

\renewcommand{\figureautorefname}{Fig.}
\renewcommand{\tableautorefname}{Table}
\renewcommand{\sectionautorefname}{Section}
\renewcommand{\subsectionautorefname}{Section}
\renewcommand{\subsubsectionautorefname}{Section}


\renewcommand{\bold}[1]{#1}
\renewcommand{\boldsymbol}[1]{#1}
\newcommand{\diag}[1]{\text{diag}(#1)}
\newcommand{\bolds}[1]{\boldsymbol{#1}}
\newcommand{\ur}{\bold{u}}
\newcommand{\up}{\bold{u}}
\newcommand{\um}{\bold{u}}
\renewcommand{\u}{\bold{u}}

\newcommand{\LamAL}{\hat{\bold{v}}}
\newcommand{\Lamm}{\bold{v}}
\newcommand{\Lamdh}{\tilde{\boldsymbol{\gamma}}}
\newcommand{\LamALn}{{\boldsymbol{\varepsilon}}}
\newcommand{\Lam}{{\boldsymbol{\lambda}}}
\newcommand{\Lamd}{{\boldsymbol{\gamma}}}
\newcommand{\scs}{\boldsymbol{\varphi}}
\newcommand{\scsd}{\boldsymbol{\psi}}
\newcommand{\btau}{\boldsymbol{\tau}}
\newcommand{\regp}{\mu}
\newcommand{\dbp}{\delta\bold{b}}
\newcommand{\dbm}{\delta\bold{b}}

\newcommand{\eref}[1]{eq. \eqref{#1}}                  
\newcommand{\fref}[1]{Figure \ref{#1}}                  

\newcommand{\tcm}{}

\newcommand{\bA}{\mathbf{A}}
\newcommand{\ba}{\mathbf{a}}
\newcommand{\bB}{\mathbf{B}}
\newcommand{\bb}{\mathbf{b}}
\newcommand{\bC}{\mathbf{C}}
\newcommand{\bD}{\mathbf{D}}
\newcommand{\bd}{\mathbf{d}}
\newcommand{\bF}{\mathbf{F}}
\newcommand{\bG}{\mathbf{G}}
\newcommand{\bg}{\mathbf{g}}
\newcommand{\bQ}{\mathbf{Q}}

\newcommand{\bI}{\mathbf{I}}
\newcommand{\bL}{\mathbf{L}}
\newcommand{\bS}{\mathbf{S}}
\newcommand{\be}{\mathbf{e}}
\newcommand{\bh}{\mathbf{h}}
\newcommand{\by}{\mathbf{y}}
\newcommand{\bR}{\mathbf{R}}
\newcommand{\br}{\mathbf{r}}
\newcommand{\bK}{\mathbf{K}}
\newcommand{\bM}{\mathbf{M}}
\newcommand{\bm}{\mathbf{m}}
\newcommand{\bP}{\mathbf{P}}
\newcommand{\bp}{\mathbf{p}}
\newcommand{\bq}{\mathbf{q}}
\newcommand{\bW}{\mathbf{W}}
\newcommand{\bU}{\mathbf{U}}
\newcommand{\bu}{\mathbf{u}}
\newcommand{\bV}{\mathbf{V}}
\newcommand{\bv}{\mathbf{v}}
\newcommand{\bZ}{\mathbf{Z}}
\newcommand{\bz}{\mathbf{z}}
\newcommand{\bX}{\mathbf{X}}
\newcommand{\bT}{\mathbf{T}}
\newcommand{\bt}{\mathbf{t}}
\newcommand{\bJ}{\mathbf{J}}
\newcommand{\Bs}[1]{\boldsymbol{#1}}
\newcommand{\B}[1]{\bold{#1}}
\newcommand{\dobs}{\tau_i^\text{obs}}



\shorttitle{Distributed proximal SVGD for Bayesian traveltime tomography}


\title [mode = title]{
Distributed Proximal Stein Variational Gradient Descent Algorithm for Large-scale Bayesian Inference in Traveltime Tomography
}  




%

\author[1]{Akshay Vishwakarma}[orcid]
\ead{akshay@igf.edu.pl}
\credit{Methodology, Software, Validation, Investigation, Visualization, Writing--original draft}

\author[1]{Kamal Aghazade}[orcid=0000-0002-2080-9697]
\ead{aghazade.kamal@igf.edu.pl}
\credit{Software, Validation, Investigation, Visualization}

\author[2]{Ali Siahkoohi}[orcid=0000-0001-8779-2247]
\ead{alisk@ucf.edu}
\credit{Methodology, Conceptualization, Software, Formal analysis, Writing--review \& editing}

\author[1]{Ali Gholami}[orcid=0000-0001-8403-1434]
\ead{agholami@igf.edu.pl}
\cortext[1]{Corresponding author}
\cormark[1]
\credit{Methodology, Conceptualization, Software, Validation, Supervision, Writing--review \& editing}

\affiliation[1]{organization={Institute of Geophysics, Polish Academy of Sciences},
            city={Warsaw},
            country={Poland}}

\affiliation[2]{organization={Department of Computer Science, University of Central Florida},
            city={Orlando},
            country={USA}}
            
%

\begin{abstract}
We present a distributed framework for large-scale Bayesian inverse problems governed by the eikonal equation, with a specific focus on seismic traveltime tomography. Traditional deterministic approaches often fail to provide the uncertainty quantification (UQ) necessary for ill-posed problems, while conventional Bayesian sampling methods such as Markov chain Monte Carlo (MCMC) suffer from the curse of dimensionality and slow convergence in high-dimensional model spaces. The proposed framework addresses these challenges through a three-tier computational strategy. First, we utilize the Fast Marching Method (FMM) to solve the eikonal equation, ensuring high numerical accuracy. Second, we reformulate the global tomographic objective into a decentralized consensus form, allowing the inversion to be decomposed into independent subproblems solved in parallel via the Alternating Direction Method of Multipliers (ADMM). This architecture eliminates the need for the explicit construction of large-scale sensitivity matrices, significantly reducing the memory footprint for 3D surveys.
Finally, we integrate Stein Variational Gradient Descent (SVGD) within the ADMM workers to perform approximate posterior sampling. By evolving a set of model particles along a functional gradient direction that balances data-fitting forces with a repulsive kernel-based diversity force, we obtain an ensemble from which posterior summaries are computed. We derive a data-space Gauss-Newton update using the Woodbury matrix identity to further accelerate the particle evolution in large-scale 3D problems. Numerical experiments on complex 2D and 3D models demonstrate that the algorithm achieves stable convergence, produces high-fidelity velocity reconstructions, and provides posterior uncertainty maps. 
\end{abstract}





\maketitle

\section{Introduction}\label{Intro}
The solution of high-dimensional inverse problems governed by non-linear partial differential equations (PDEs) is a cornerstone of modern computational science. In these systems, characterizing the inherent ill-posedness and solution non-uniqueness is essential for reliable scientific inference. While deterministic optimization methods seek a single best-fit model, Bayesian inference provides a principled framework for uncertainty quantification (UQ) by representing the solution as a posterior probability distribution conditioned on observed data \citep{Tarantola_2005_IPT,Jin_2010_HBI}. 
Traveltime tomography has a wide range of applications spanning atmospheric science, medical imaging, and geophysics. These applications include the reconstruction of internal temperature and flow fields from boundary measurements \citep{Othmani_2023_ATR,Kira_2025_FWA}, as well as imaging the internal structure of the Earth using seismic wave traveltimes recorded at the Earth's surface \citep{Nolet_2008_BST,KLIBANOV_2023_NS3D,Loris_2010_NRT,Fichtner_2024_ST,Si_2025_HoAA}.


The primary advantage of traveltime-based methods lies in their computational robustness and efficiency compared to wave-equation-based approaches such as full-waveform inversion (FWI) \citep{ Gao_2025_LATTE}. 
Energy propagation in this context is governed by the eikonal equation, a non-linear first-order PDE representing the high-frequency approximation of the wave equation. Standard grid-based solvers, such as the Fast Marching Method (FMM), can be applied to efficiently solve the forward problem with $O(N\log N)$ computational complexity  \citep{Si_2025_HoAA}. 
Using fast sweeping methods, the problem can be solved with a complexity of $O(N)$
\citep{Zhao_2005_FSM,Fomel_2009_FSM}. 
While FWI exploits the full information content of seismograms, it is computationally intensive, highly non-linear, and prone to falling into local minima when an accurate initial model is unavailable \citep{Virieux_2009_OFW,Operto_2023_ESS}. In contrast, traveltime tomography is less sensitive to the starting model and can yield stable low-wavenumber models even from low-quality waveforms or sparse data sampling \citep{Bishop_1985_TDV,Zhang_1998_NRT}. Such models provide the essential long-wavelength background model required for the success of high-resolution imaging workflows, such as depth migration and FWI \citep{Virieux_2009_OFW,Operto_2023_ESS}.


Despite these advantages, some challenges still remain, specifically, in the inversion part.
The primary bottleneck in modern large-scale tomography arises from the prohibitive computational and memory requirements associated with high-dimensional model spaces. For large-scale 3D seismic surveys, the explicit construction and storage of the Gauss-Newton Hessian matrix becomes computationally and memory-prohibitive. While the adjoint-state method \cite{Si_2025_HoAA} has mitigated some memory limitations by allowing for the direct computation of the misfit gradient, the inversion process remains computationally intensive, requiring numerous iterations of forward and adjoint modeling proportional to the number of seismic sources. These challenges are further exacerbated when seeking a Bayesian solution. Recent efforts in distributed tomography have explored various architectures to enhance scalability in deterministic settings \citep{Shin_2022_DTT}. 

Traveltime tomography is inherently ill-posed and underdetermined. Instability often arises from the clustering of small eigenvalues in the sensitivity operator and the presence of random noise in recorded data \citep{Gholami_2010_RLN,Loris_2010_NRT}. The inherent instability of such inverse problems necessitates the integration of prior information through regularization to ensure stable and physically plausible solutions.
 While traditional Tikhonov-type $\ell_2$ regularization \citep{Tikhonov_1977_SIP} is widely employed for its computational simplicity, it typically produces biased, over-smoothed reconstructions that smear out sharp structural interfaces and high-resolution details \citep{KLIBANOV_2023_NS3D}. To achieve high-fidelity results, non-smooth transform-domain sparsity-promoting regularizers such as Total Variation (TV) \citep{Rudin_1992_NTV,Gholami_2010_RLN, Loris_2010_NRT,Burgel_2017_ASR}, have become essential for their unique capacity to promote piecewise-smooth model features while preserving sharp physical discontinuities.
Incorporation of these functionals introduces the challenge of losing smoothness at certain points. The proximal mappings and variable-splitting methods address this difficulty by decoupling the differentiable data-misfit subproblems from the non-smooth regularization tasks \citep{Goldstein_2009_SBM}.
 

Another challenge in most current tomographic methods is the lack of uncertainty quantification (UQ) \citep{Ryberg_2018_BIR}. Relying on a single deterministic model estimate ignores the inherent non-uniqueness of the inverse problem and may lead to overconfident interpretations of the results. Bayesian inference provides a principled framework to address this by representing the solution as a posterior probability distribution conditioned on observed data. However, generating samples from the posterior in high-dimensional model spaces remains a major computational hurdle, particularly when constrained by the non-linear physics of wave propagation.

Classical Markov chain Monte Carlo (MCMC) methods \citep{Mosegaard_1995_MCS,Martin_2012_SNM} are asymptotically exact but inherently sequential, and their mixing degrades in high dimensions \citep{Curtis_2001_PIS,Zhang_2023_3DB}, rendering them prohibitive at the scale of modern multi-source tomography, where each sample requires one eikonal solve per source. Parametric variational methods \citep{Rizzuti_2020_PUD,Zhao_2024_PSV,Zhao_2025_EBF} sidestep the sequential bottleneck by fitting a trainable family to the posterior, but expose a tradeoff between flexibility and tractability: simple choices such as mean-field Gaussians systematically underestimate uncertainty \citep{Zhang_2023_3DB}, while expressive alternatives such as normalizing flows \citep{Zhao_2022_BST,Yin_2025_MVI} require many trainable parameters and face their own optimization challenges. Stein variational gradient descent (SVGD) \citep {Liu_2016_SVGD} avoids these parametric restrictions altogether by representing the posterior as an ensemble of particles that it transports toward the target by combining log-posterior gradients with a kernel-based repulsive term, and has shown promise for Bayesian seismic imaging and tomography \citep{Zhang_2020_VFW,Zhang_2023_3DB, Corrales_2025_ASV}. A critical obstacle nonetheless arises: SVGD cannot handle the non-smooth regularizers (TV or $\ell_1$ sparsity) that stabilize ill-conditioned tomography, since their log-priors have undefined gradients at the kinks and cannot enter the particle update directly. While derivative-free particle methods such as ensemble Kalman inversion \citep{Li_2025_STT}, which recasts the inverse problem as an artificial dynamical system and evolves an initial ensemble through Kalman-filter updates driven by its empirical covariance, sidestep this differentiability requirement, they are fundamentally restricted to Gaussian posteriors and therefore cannot capture the multimodality or heavy tails that a Bayesian treatment is meant to reveal. To resolve this challenge, we follow the dual-space framework of \citet{Siahkoohi_2026_DSP}, which enables SVGD to sample from otherwise intractable constrained posteriors by relaxing the constraints into auxiliary-variable splits in an augmented Lagrangian (AL) and progressively enforcing them through multiplier updates. We extend this mechanism to both the data constraints and the non-smooth prior: a consensus splitting decouples the per-source data terms, while the prior enters only through a proximal operator, yielding an entirely smooth target distribution amenable to SVGD.

In this work, we propose a distributed proximal SVGD algorithm that achieves computational scalability through a variable splitting framework. By reformulating the global tomographic objective into a decentralized consensus form, we utilize the Alternating Direction Method of Multipliers (ADMM)  to decompose the multi-source problem into independent subproblems solved in parallel \citep{Boyd_2011_DOS,Tsianos_2012_CBD,Zand_2020_COT}. This architecture enables the particle ensemble to sample from posterior distributions constrained by non-differentiable priors while preserving sharp interfaces and edge-enhancing properties. The primary contributions of this paper are: 
\begin{enumerate}[i)]
\item The formulation of a distributed proximal SVGD framework for high-dimensional Bayesian inference with non-smooth regularization constraints.

\item The derivation of a data-space Gauss--Newton update using the Woodbury matrix identity, enabling efficient computation of the Gauss--Newton step in data space while significantly reducing the dimensionality of particle evolution.

\item The development of a parallelization strategy that decomposes the multi-source, multi-particle inversion problem into independent source- and particle-specific subproblems. Specifically, for $N_s$ sources and $N_p$ particles, the framework solves $N_s \times N_p$ mutually independent local subproblems concurrently.

\item An evaluation of the proposed framework in terms of parallel scalability, inversion accuracy, and uncertainty quantification performance on challenging 2D and 3D benchmark problems, including calibration of the particle ensemble against a converged MCMC reference on a 1D problem small enough for the latter to be affordable.


\end{enumerate}

The remainder of this paper is organized as follows. In \Cref{sec_theory}, we establish the theoretical foundation of the forward problem, detailing the eikonal equation and its numerical solution using the FMM.~\Cref{sec_deterministicInv} presents our deterministic inversion framework, where the tomographic objective is reformulated into a decentralized consensus form and solved via the ADMM to achieve computational scalability. In \Cref{sec:bayes}, we extend this architecture to a fully Bayesian setting by integrating a distributed proximal SVGD algorithm, enabling efficient posterior sampling and UQ in high-dimensional spaces.~\Cref{sec:results} evaluates the performance of the proposed framework through a series of numerical experiments on 1D and complex 2D and 3D benchmarks, focusing on reconstruction fidelity, parallel efficiency, and the quality of standard deviation maps. Finally, concluding remarks and directions for future research are provided in \Cref{sec_summary}.

\section{Theory} \label{sec_theory}

For an isotropic medium characterized by a spatially varying slowness model $m(\bx)$ with $\bx=(x, y, z)$, the traveltime field $T(\bx;\bx_s)$ corresponding to a point source located at $\bx_s=(x_s, y_s, z_s)$ is governed by the eikonal equation, which represents the high-frequency (ray-theoretical) approximation of the wave equation \citep{Nolet_2008_BST}:
\begin{equation} \label{eikonal}
|\nabla T(\bx;\bx_s)|^2 = m(\bx)^2 = \frac{1}{v^2(\bx)}, \quad \forall \bx \in \Omega,
\end{equation}
subject to the boundary condition
\begin{equation*}
T(\bx_s;\bx_s) = 0.
\end{equation*}
Here, $\nabla$ denotes the spatial gradient operator, $|\cdot|$ represents the magnitude, and $v(\bx)$ is the velocity at position $\bx$. The computational domain is denoted by $\Omega \subset \mathbb{R}^d$, with $d=2$ or $3$.

For each source $\bx_s^i, i=1,...,N_s$,
and receivers $\bx_r^l,\ l=1,\dots,N_r^i$, the forward operator $G_i: \mathbb{R}^N \to \mathbb{R}^{N_r^i}$ collects the predicted traveltimes, 
\begin{equation*}
[G_i(m)]_{l}=T(\bx_r^l;\bx_s^i),\qquad l=1,\dots,N_r^i,
\end{equation*}
where $N_r^i$ is the number of receivers associated with source $\bx_s^i$. The corresponding measured data are stored in the column vector $\dobs \in \mathbb{R}^{N_r^i}$. 
\subsection{Model Discretization}
\medskip
To solve the forward problem for arbitrary source and receiver geometries, the subsurface is parameterized on a regular Cartesian grid. The velocity model $v(\bx)$ is discretized as a vector $v \in \mathbb{R}^N$, where $N = N_x \times N_y \times N_z$ denotes the total number of grid nodes, and $N_x$, $N_y$, and $N_z$ are the number of nodes along each spatial dimension. The corresponding slowness model is defined componentwise by $m_j=1/v_j,\, j=1,\dots, N$.
Because source and receiver locations generally do not coincide with grid nodes, interpolation operators are required. In this work, we employ linear interpolation to map quantities between the continuous physical domain and the discrete computational grid.

\subsubsection{Fast Marching Method and Calculation of Sensitivities}\label{sec:FMM}

We solve the eikonal equation \eqref{eikonal} using the second-order FMM, a grid-based numerical algorithm designed to compute first-arrival traveltimes efficiently \citep{Patrick_2011_FM,Sethian_1999_FMM,Treister_2016_FMA}. The FMM propagates a monotonically advancing wavefront across the computational domain, ensuring causality and unconditional stability.
Compared to traditional two-point ray-tracing methods, which may suffer from convergence issues in complex, heterogeneous media, the FMM is highly robust and reliably computes globally minimal traveltimes, even in the presence of strong velocity contrasts and shadow zones.

For each source location $\bx_s$, the eikonal equation is solved on the discretized grid to obtain the traveltime field $T(\bx;\bx_s)$. This procedure guarantees that the computed traveltimes correspond to the first-arrival solution, thereby avoiding the local minima issues commonly encountered in shooting or bending ray-tracing approaches \citep{Nolet_2008_BST}.
Once the traveltime field is available, ray paths can be recovered, if needed, by backtracking from receiver locations to the source $\bx_s$ along the negative gradient of the traveltime field, i.e., $-\nabla T(\bx;\bx_s)$. The sensitivity (Jacobian) matrix is constructed by integrating the path length of each ray through the model grid. Specifically, the contribution of each ray segment is accumulated along its trajectory and distributed to the neighboring grid nodes using linear interpolation.

\begin{figure}
    \centering
    \includegraphics[width=\linewidth, trim={0cm 0.5cm 0cm 1.25cm},clip
]{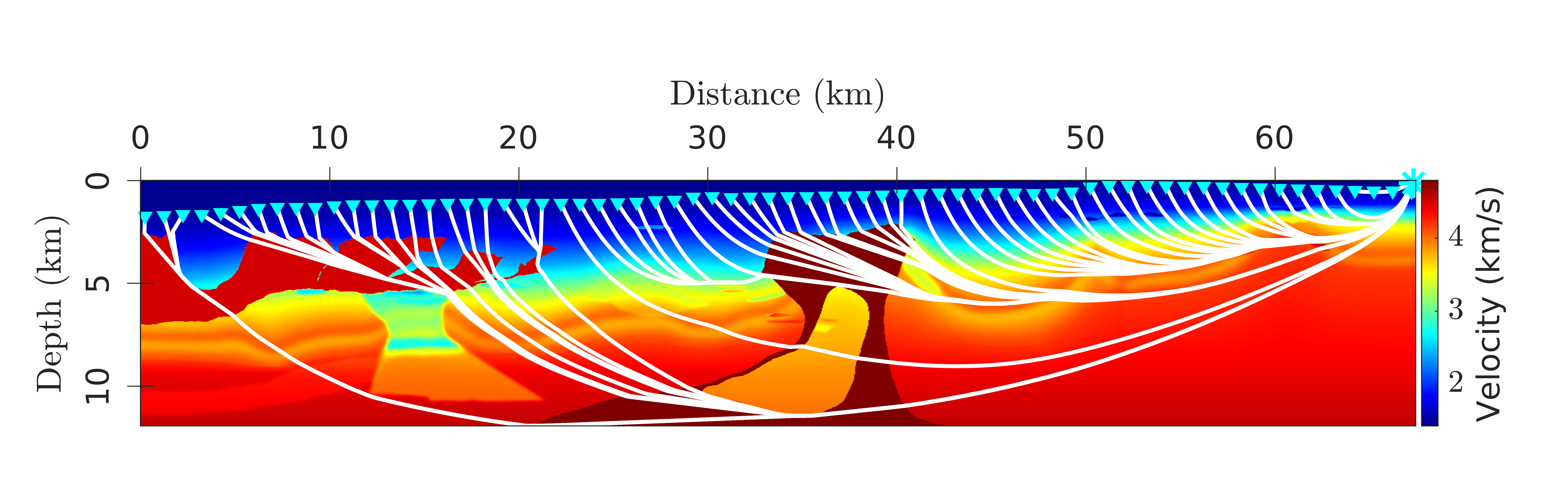}
    \includegraphics[width=\linewidth, trim={0cm 0.5cm 0cm 1.25cm},clip
]{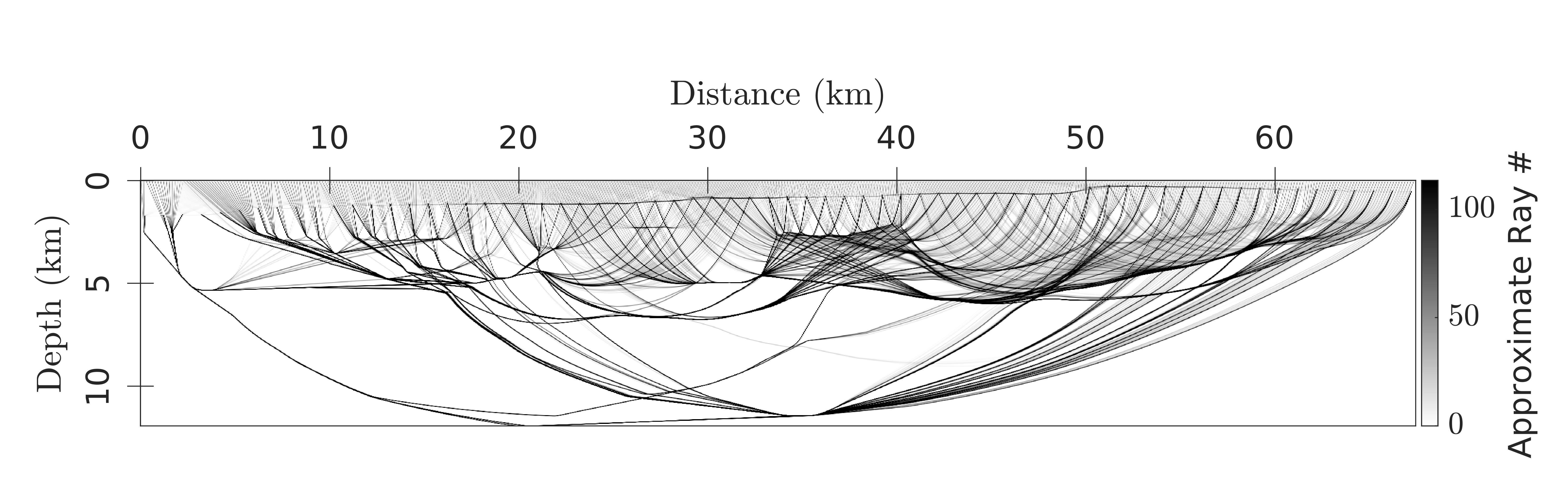}
    \caption{Acquisition geometry and illumination for the BP 2004 salt benchmark.
    Top: True velocity model. Receivers (cyan triangles) are deployed along the seafloor at a regular spacing of 1 km, while sources are located at the sea surface. The white curves represent geodesic rays connecting a source located at the top-right corner of the model to each receiver. Bottom: The dimensionless illumination map generated by summing the sensitivity kernels across all source-receiver pairs. Each grid node is normalized by the horizontal grid spacing $\Delta x$, representing the approximate number of rays per cell. Regions of high intensity (darker colors) indicate well-constrained areas where the subsurface structure is heavily sampled by crossing ray paths.}
    \label{fig:BP_rays}
\end{figure}

\section{Deterministic Inversion Methodology} \label{sec_deterministicInv}

Seismic traveltime tomography is typically formulated as an inverse problem that seeks to minimize the discrepancy between observed arrival times and those predicted by a velocity model. When sources are positioned outside the domain of interest and receivers surround the domain (full-boundary acquisition), \citet{KLIBANOV_2023_NS3D} demonstrated that the traveltime inverse problem can be convexified using Carleman-weighted formulations. In this study, however, we consider a more challenging and realistic acquisition geometry commonly encountered in seismic exploration.
Specifically, both sources and receivers are located only at the surface, resulting in incomplete illumination of the subsurface. This acquisition geometry creates poorly constrained regions that are not traversed by ray paths \citep{Nolet_2008_BST,Loris_2010_NRT,Fichtner_2024_ST,Si_2025_HoAA,Gholami_2010_RLN}. Consequently, portions of the model belong to the null space of the forward operator, making regularization essential for obtaining stable and physically meaningful solutions \citep{Tarantola_2005_IPT}.

An example of this challenge is illustrated in \cref{fig:BP_rays} using the 2004 BP salt velocity benchmark model \citep{billette20052004}. The velocity model contains sedimentary basins characterized by progressively increasing velocity with depth, as well as high-velocity, high-contrast salt bodies. The model extends over a domain of 67.5 km $\times$ 12 km and is discretized with a grid spacing of 25 m. We consider an ultra-long-offset fixed-spread ocean-bottom seismometer (OBS) acquisition geometry consisting of $N_r=67$ receivers spaced 1 km apart and $N_s=450$ pressure sources distributed uniformly every 150 m at a depth of 25 m below the sea surface.

\Cref{fig:BP_rays} (top panel) shows the geodesic ray paths connecting a source located at the top-right corner of the model to each receiver (cyan triangles). The bottom panel shows the illumination map (ray density) across the domain for all sources and receivers. As can be seen, the sparse acquisition geometry results in limited ray coverage of the subsurface. Large regions remain weakly illuminated or entirely unsampled, highlighting the ill-posed nature of the inversion problem and the necessity of incorporating appropriate regularization.

\subsection{Global Objective Function}
For a survey involving $N_s$ seismic sources, the global inverse problem can be formulated as the following constrained optimization problem:
\begin{equation}
    \minimize_{m}~\mathcal{R}(Lm)\quad \text{subject to}\quad G_i(m)=\dobs,\quad i=1,...,N_s,
\end{equation}
where $\mathcal{R}(Lm)$ is a global regularization functional and $G_i(m)$ is the nonlinear forward operator mapping the model parameters to predicted traveltimes for the $i$-th source. 
In this formulation, the model $m$ is assumed to exhibit certain structural or statistical properties when transformed by a linear analysis operator $L\in \mathbb{R}^{N'\times N}$, referred to as the regularization operator. The rows of $L$ typically correspond to basis functions or differential operators, such as wavelets or finite-difference approximations of spatial derivatives (e.g., in Tikhonov or TV regularization) \citep{Aster_2018_PEI,Loris_2007_TIU,Gholami_2010_RLN}. In many cases, $L$ is not invertible, particularly when it represents a derivative operator.

In a general form, the regularization term can be expressed as
\begin{equation*}
\mathcal{R}(Lm) = \sum_{i} \phi\big( [Lm]_i \big),
\end{equation*}
where $\phi(\cdot)$ is a potential function that reflects the assumed prior distribution of the transformed coefficients $Lm$. For example, a quadratic potential, $\phi(x)=x^2$, corresponds to Gaussian priors (Tikhonov regularization), while sparsity-promoting choices such as $\ell_1$ norms, $\phi(x)=|x|$, correspond to Laplace-type priors \citep{Tarantola_2005_IPT}.

This constrained problem can be solved using the method of multipliers  
which solves the following saddle-point problem:
\begin{equation} \label{min_max}
    \minimize_{m}\maximize_{\lambda_1,...,\lambda_{N_s}}~\mathcal{R}(Lm)+ \sum_{i=1}^{N_s} \langle \lambda_i , G_i(m)-\dobs \rangle+ \sum_{i=1}^{N_s}\frac{\mu}{2} \|G_i(m)-\dobs\|_2^2,
\end{equation}
where $\lambda_i$ denotes the Lagrange multiplier (dual variable) associated with the $i$-th source and $\mu > 0$ is the penalty parameter. 

To solve the min-max problem in \eqref{min_max}, we use the AL algorithm discussed in detail, e.g., in \citep{Powell_1969_NLC,Nocedal_2006_NO,Boyd_2011_DOS}. Introducing the scaled dual variable $\lambda_i\leftarrow \frac{1}{\mu} \lambda_i$, the AL method alternates between the following two steps until a convergence criterion is satisfied:
\begin{enumerate}
    \item With the multiplier fixed at its current value $\lambda_i^k$, minimize the AL objective with respect to $m$ to obtain the updated model:
    \begin{equation} \label{AL_p}
    m^{k+1}=\argmin_{m}~\sum_{i=1}^{N_s}\frac{\mu}{2}\|G_i(m)-\dobs+\lambda_i^k\|_2^2+ \mathcal{R}(Lm),
\end{equation}
\item With the model fixed at its updated value $m^{k+1}$, perform a gradient-ascent step on the AL objective with respect to $\lambda$:
\begin{equation}\label{AL_d}
    \lambda_i^{k+1} =\lambda_i^k + G_i(m^{k+1})-\dobs,~i=1,...,N_s.
\end{equation} 
\end{enumerate}
%
While the unconstrained formulation in \eqref{AL_p} provides a straightforward path for model recovery in the presence of noisy data (by selecting a proper value for $\mu$), we adopt the constrained formulation and its associated method of multipliers primarily to leverage a dual-grid parameterization strategy that decouples the inversion and forward modeling requirements. In this framework, the primal update in \eqref{AL_p} is performed on a coarser inversion grid, which is essential for maintaining computational efficiency and aligning the model's degrees of freedom with the limited resolving power of the traveltime data. Conversely, the dual update in \eqref{AL_d} and the underlying eikonal forward solver utilize a finer grid to guarantee high numerical accuracy and to honor the complex physics of wave propagation. The critical advantage of this constrained approach is the error-correction role played by the Lagrange multipliers, which accumulate the traveltime residuals at each iteration. These multipliers effectively steer the coarse-grid model updates toward feasibility, compensating for the discretization errors inherent in the coarser mesh while allowing the inversion to benefit from the reduced dimensionality and enhanced stability of a lower-parameter space.

The description for each step is as follows:
\begin{itemize}
\item Primal update step \cref{AL_p}:  
When $\boldsymbol{\lambda}_i^k = 0$, this subproblem reduces to the classical multi-source traveltime tomography formulation \citep{Nolet_2008_BST}. However, solving this problem presents several challenges: 
\begin{enumerate}[i)]
    \item For effective regularization strategies such as $\ell_1$-sparsity or TV, the regularization functional $\mathcal{R}$ is non-differentiable. Consequently, specialized nonsmooth optimization methods (e.g., proximal algorithms) are required \citep{Gholami_2010_RLN,Loris_2010_NRT}. 
    \item Even when $\mathcal{R}$ is smooth, the associated Gauss--Newton or full-Newton Hessian is defined in the model space and has size $N \times N$, which becomes computationally prohibitive for large-scale 3D problems \citep{Gao_2025_LATTE}.
    \item The choice of the strength of the regularization parameter is nontrivial \citep{Rawlinson_2016_OUS}. Its optimal value depends on the noise level $\sigma$ and problem scaling, and naive selection often requires multiple costly inversions, making adaptive strategies essential.
\end{enumerate}

\item Dual update step \cref{AL_d}:
This subproblem corresponds to a dual ascent step that updates the Lagrange multipliers to enforce consistency with the traveltime constraints. Specifically, the multipliers accumulate the traveltime residuals and drive the iterates toward feasibility, i.e., $G_i(m) \rightarrow \dobs$, largely independent of the choice of $\mu$ \citep{Nocedal_2006_NO}. However, the rate of convergence is affected by $\mu$.
This property plays a crucial role: it allows relatively small values of $\mu$ to be used to improve the conditioning of the primal subproblem (e.g., stabilizing the Hessian), while still ensuring that the data constraints are satisfied asymptotically through the dual updates.  
The method of multipliers balances data fitting and regularization through the interaction between primal and dual updates. While the primal step enforces regularization and model smoothness, the dual step progressively corrects constraint violations. Under suitable conditions, this interplay leads to convergence toward a solution that satisfies both the data constraints and the imposed prior \citep{Gholami_2024_FWI}.
\end{itemize}

Although the classical formulation remains computationally demanding due to the large-scale nature of the primal subproblem, it provides a natural foundation for distributed and consensus-based extensions. In particular, by introducing auxiliary variables, the problem can be decomposed into smaller source-wise subproblems, enabling parallelization and improved scalability.

Generally, it is not necessary to solve the primal subproblem \cref{AL_p} to full convergence to achieve the convergence of the whole algorithm. Following a diagonalized variant of algorithm \citep{Miele_1972_OTM}, we perform only a single iteration of a Gauss-Newton method to partially solve the subproblem in~\cref{AL_p}~ and hence increase the efficiency while maintaining the desirable properties of the algorithm. 
We use the first-order approximation,
\begin{equation*}
    G_i(m) \approx G_i(m^k) + J_i^k [m-m^k],
\end{equation*}
where $J_i^k$ is the sensitivity matrix (or Jacobian) defined by
\begin{equation*}
     J_i^k =\left. \frac{\partial G_i(m)}{\partial m}\right\vert_{m=m^k}.
\end{equation*}
This allows us to linearize the primal subproblem in \cref{AL_p}:
\begin{equation}\label{AL_p_lin}
m^{k+1}=\argmin_{m}~\sum_{i=1}^{N_s}\frac{\mu}{2}\|J_i^k m -\dobs + d_i^k \|_2^2+ \mathcal{R}(Lm),
\end{equation} 
where $d_i^k=G_i(m^k)-J_i^k m^k+\lambda_i^k$.
\subsection{Consensus Optimization and Variable Splitting}
\medskip
To solve the large-scale optimization problem in \cref{AL_p_lin} efficiently in distributed computing environments, we reformulate it into a global variable consensus form. We introduce $N_s$ local copies of the model parameters, $p_i$, each associated with an individual source $i$. These local variables are constrained to match a unique global consensus model $m$. In addition, we introduce an auxiliary variable $u = Lm$ to decouple the data misfit term from the regularization term. This variable splitting separates the smooth quadratic data-fitting component from the potentially non-smooth regularization functional, enabling the use of efficient proximal algorithms \citep{Goldstein_2009_SBM, Combettes_2011_PRO}.
This reformulation changes the primal subproblem \eqref{AL_p_lin} to the following constrained optimization problem:
\begin{mini}
{\{p_i\}, m,u}{\sum_{i=1}^{N_s}\frac{\mu}{2}\|J_i^k p_i-\dobs+d_i^k\|_2^2+  \mathcal{R}(u)}{}{}
\addConstraint{p_i-m}{=0 ,\quad}{ i =1 ,\ldots ,N_s}
\addConstraint{u-Lm }{=0 ,}{}
\label{consus}
\end{mini}
The AL function associated with \eqref{consus} is
\begin{equation*} \label{AL}
\mathcal{L}(\{p_i\}, \{q_i\}, m,u,v)= \sum_{i=1}^{N_s}\left\{\frac{\mu}{2}\|J_i^k p_i-\dobs+d_i^k\|_2^2
+ q_i^{\top}(p_i-m) + \frac{\alpha}{2}\|p_i-m\|_2^2\right\}
+ v^{\top}(u-Lm) + \frac{\beta}{2}\|u-Lm\|_2^2 + \mathcal{R}(u) 
\end{equation*}
where $q_i \in \mathbb{R}^N,\, i=1,...,N_s$ and $v$ are the dual variables associated with the respective constraints, and $\alpha,~\beta>0$ are the penalty parameters.
The optimization problem is
\begin{equation}
    \minimize_{\{p_i\},\, m,\, u}\maximize_{\{q_i\},\, v } \mathcal{L}(\{p_i\}, \{q_i\}, m,u,v).
\end{equation}
This optimization problem can be solved efficiently using the scaled form ADMM \citep{Boyd_2011_DOS}, which decomposes the global optimization problem into a sequence of smaller subproblems associated with different parameter blocks. These subproblems are then solved iteratively in an alternating manner.
In the following, we describe the solution strategy for each subproblem in detail.

\begin{enumerate}
    \item Local model update (worker step): Each source $i$ independently solves a damped tomographic problem to generate a local update $p_i^{k+1}$:
    \begin{equation} \label{local}
    p_i^{k+1} = \argmin_{p_i} \left(\frac{\mu}{2}\|J_i^k p_i-\dobs+d_i^k\|_2^2 + \frac{\alpha}{2} \|p_i - m^k + q_i^k\|_2^2 \right).
    \end{equation}
    This update forces the local model to fit the $i$-th source data while remaining in the vicinity of the current global model.
The optimality condition for the least-squares problem \eqref{local} leads to the following closed-form expression for the local image $p_i^{k+1}$: 
\begin{align} \label{pi_k}
 p_i^{k+1} 
  &=m^k + \left((J_i^k)^{\top} J_i^k + (\alpha/\mu) I_N\right)^{-1}\left[(J_i^k)^{\top} (\dobs-G_i(m^k)-\lambda_i^k )- (\alpha/\mu) q_i^k\right].
\end{align}
Using the following Sherman–Morrison–Woodbury (SMW) matrix identity (see~\Cref{appen_Woodbury}):
\begin{equation} \label{eq:smw}
\left((J_i^k)^{\top}J_i^k + (\alpha/\mu) I_{N} \right)^{-1}=
\frac{\mu}{\alpha}I_{N} - \frac{\mu}{\alpha}(J_i^k)^{\top}
\left(J_i^k(J_i^k)^{\top} +  (\alpha/\mu) I_{N_r^i} \right)^{-1} J_i^k,
\end{equation}
gives the following data-space formulation:
\begin{equation} \label{local_p}
p_i^{k+1} = m^k 
-\frac{\mu}{\alpha}g_i^k + \frac{\mu}{\alpha}(J_i^k)^{\top}
\left(J_i^k(J_i^k)^{\top} +  (\alpha/\mu) I_{N_r^i} \right)^{-1} J_i^k g_i^k,
\end{equation}
where $g_i^k=(J_i^k)^{\top} (G_i(m^k)-\dobs+\lambda_i^k)+ (\alpha/\mu)q_i^k$.
\medskip 
In this formulation, we only need to invert the data-space Hessian matrix $J_i^k(J_i^k)^{\top} +  (\alpha/\mu) I_{N_r^i}$, which is of size $N_r^i\times N_r^i$.
 \item Global consensus update (master step): The master node collects the local updates and computes a unique global model $m^{k+1}$ by solving:
\begin{equation} \label{m_subprob}
    m^{k+1} =\argmin_{m} \sum_{i=1}^{N_s} \frac{\alpha}{2} \|p_i^{k+1} - m + q_i^k\|_2^2+
\frac{\beta}{2} \|u^k - Lm + v^k\|_2^2.
\end{equation}
This problem is quadratic in $m$ and admits the following closed-form solution obtained from the normal equations: 
\begin{equation}
    \left(\alpha N_s I_N + \beta L^{\top}L\right) m^{k+1}=\sum_{i=1}^{N_s} \alpha (p_i^{k+1}+ q_i^k)+
\beta L^{\top}(u^k + v^k).
\end{equation}
The resulting linear system corresponds to a Tikhonov-regularized least-squares problem. When $L$ represents a convolutional operator (e.g., finite-difference gradients), the system matrix $\alpha N_s I_N + \beta L^{\top} L$ is diagonalizable in the Fourier or DCT domain, allowing the update to be computed efficiently using Fast Fourier Transform (FFT)-based solvers.
    

    \item Global dual update: The auxiliary variable $u^{k+1}$ is updated via a proximal mapping:
    \begin{equation}
    u^{k+1} = \text{prox}_{\frac{\mathcal{R}}{\beta}} \left( Lm^{k+1} - v^k \right) = \argmin_{u} \left( \mathcal{R}(u) + \frac{ \beta}{2} \|u - Lm^{k+1} + v^k\|_2^2 \right),
    \end{equation}
    This step corresponds to the proximal operator associated with the regularization functional $\mathcal{R}$. It can be computed efficiently and, for many commonly used regularizers (e.g., $\ell_1$-norm or TV), admits closed-form or highly efficient numerical solutions \citep{Combettes_2011_PRO}.
    
    \item Dual variable update: Each worker updates its dual variable to enforce the consensus constraint in future iterations:
\begin{align}
    q_i^{k+1} &= q_i^k+p_i^{k+1} - m^{k+1},~i=1,...,N_s,\\
    v^{k+1} &= v^k+u^{k+1} - Lm^{k+1}.
    \end{align}
\end{enumerate}
\Cref{alg:det} summarizes the proposed consensus ADMM framework. The method is inherently well-suited for large-scale 2D and 3D traveltime tomography problems. Its main advantages are summarized as follows:
\begin{itemize}

\item 
The local model updates are carried out independently by a set of parallel workers, each associated with a source. This decomposition reduces the computational burden of handling large Jacobian matrices, as each worker operates on a data-space Hessian whose size scales with the number of receivers rather than the number of model parameters. Moreover, the Jacobian matrices are sparse and can be computed and applied efficiently even in large-scale 3D settings.

\item 
The global consensus update is formulated as a structured linear system that can be solved efficiently using FFT-based methods.

\item 
Non-smooth regularization terms are naturally incorporated through proximal operators. This enables the use of advanced regularizers, such as TV, without compromising computational efficiency.

\end{itemize}

\subsection{Penalty Parameter Selection and Model Initialization}
\medskip
\subsubsection{Adaptive Selection of Penalty Parameters}
\medskip
The local model update $p_i^{k+1}$ defined in \cref{local_p} corresponds to the solution of the linearized forward problem and depends explicitly on the penalty parameter ratio $\gamma=\alpha/\mu$. 
A simple and effective rule selects $\gamma$ separately at every iteration and every source, 
\begin{equation}
    \gamma_i^{k} = \delta \, \rho \big(J_i^k (J_i^k)^{\top} \big), \qquad \delta>0,
\end{equation}
where $\rho(\cdot)$ denotes the spectral radius, i.e., the largest eigenvalue of the data-space Gauss-Newton Hessian \citep[e.g.,][]{Loris_2010_NRT}. 
Alternatively, $\gamma$ can be selected using data-driven approaches. A straightforward derivation (\Cref{app:B}) shows that the associated traveltime residual can be written as
\begin{equation*}
    r_i(\gamma)= -\left(\frac{1}{\gamma}J_i^k (J_i^k)^{\top} + I_{N_r^i}\right)^{-1}
    \left[\dobs-G_i(m^k)-\lambda_i^k + J_i^k q_i^k\right].
\end{equation*}
%
%
Since the matrix $J_i^k (J_i^k)^{\top}$ is relatively small, its singular value decomposition (SVD) can be computed efficiently on each local worker. This enables the use of parameter-selection strategies such as generalized cross-validation (GCV) \citep{wahba1990spline} and the residual whiteness principle (RWP) \citep{lanza2020residual,almeida2013parameter,aghazade2025automatic} to determine an appropriate value of $\gamma$.
 
\subsubsection{Initial Model Construction}\label{sec:init}
\medskip
Due to the inherent nonlinearity of traveltime tomography, an initial model is required to start the inversion, and it plays a critical role in the convergence and reliability of the final solution. Because of the large null space of the forward operator, different initial models may lead to different reconstructed models that fit the observed traveltimes equally well \citep{Zhang_1998_NRR}.
A commonly used and physically reasonable choice is a laterally homogeneous velocity model with a monotonic increase in velocity with depth $z$. This type of model is generally consistent with geological settings and provides a stable starting point for the inversion.
This kind of gradient velocity model offers two main advantages:
\begin{enumerate}
\item It is fully described by only two parameters, namely the intercept $a$ and the vertical gradient $b$. Accordingly, the initial slowness model $m^0$ is given through the relation
\begin{equation*}
    m^0(z) = \frac{1}{a + b z},
\end{equation*}
where $z$ denotes depth.

\item For any source–receiver pair, the traveltime can be computed analytically, which makes this model particularly attractive for global optimization methods to search for the parameters $a$ and $b$ in a least squares sense.
The underlying principle is that, in a depth-stratified medium where velocity depends only on depth, seismic rays follow circular trajectories in the vertical plane of propagation.
For a source located at $(x_s, z_s)$ and a receiver at $(x_r, z_r)$ in a 2D problem, the analytic traveltime $\tau$ is given by \citep{Fomel_2009_FSM,Slotnick_1959_ETaP}
\begin{equation*}
\tau
=
\frac{1}{b}
\ln\!\left(
\frac{
(z_s - z_c)\,\bigl[\varrho + (x_r' - x_c)\bigr]
}{
(z_r - z_c)\,\bigl[\varrho + (x_s' - x_c)\bigr]
}
\right),
\end{equation*}
where
\[
\varrho = \sqrt{(x_s' - x_c)^2 + (z_s - z_c)^2},
\]
and the circle center is defined by
\begin{equation*}
x_c =
\frac{
(x_r')^2 - (x_s')^2
+ (z_r - z_c)^2
- (z_s - z_c)^2
}{
2(x_r' - x_s')
},
\qquad
z_c = -\frac{a}{b}.
\end{equation*}

In this formulation, $x_s'$ and $x_r'$ denote the projected horizontal coordinates of the source and receiver onto the vertical plane containing both points.
\end{enumerate}


\begin{algorithm}
\caption{Distributed Multi-source Traveltime Tomography via Consensus ADMM (Deterministic)}
\label{alg:det}
\begin{algorithmic}[1]
\State \textbf{Input:} Observed traveltimes $\{\dobs\}_{i=1}^{N_s}$, source locations $\{\mathbf{x}_{s}^i\}_{i=1}^{N_s}$, regularization $\mathcal{R}(m)$, weights $\alpha,\, \beta$
\State \textbf{Initialize:} 
\State \quad Global model $m^0$
\State \quad Local variables $p_i^0 = m^0$, global variable $u^0=0$, dual variables $v^0=0,\,\lambda_i^0=0,\,q_i^0 = 0$ for $i=1,\dots,N_s$, $k=0$

\While{not converged}

\State \textbf{Worker step (parallel over sources $i=1,\dots,N_s$):}

\State \quad Solve the eikonal equation using FMM to obtain traveltime field $T^k_i$ in $m^k$
\State \quad Compute the traveltimes  $G_i(m^{k})$ and the Jacobian $J_i^k$ 
\State \quad Compute the penalty parameter $\gamma_i^{k}$
\State \quad Compute the gradient:
\State \quad $g_i^k=(J_i^k)^{\top}(G_i(m^{k})-\dobs+\lambda_i^k)+ \gamma_i^k q_i^k$
\State \quad Local model update (data-space Gauss-Newton):
\State \qquad $\Delta p_i^k = -\frac{1}{\gamma_i^{k}} g_i^k + \frac{1}{\gamma_i^{k}} (J_i^k)^{\top} 
(J_i^k (J_i^k)^{\top} + \gamma_i^{k} I_{N_r^i})^{-1} J_i^k g_i^k$

\State \qquad $p_i^{k+1} = m^k + \Delta p_i^k$

\State \textbf{Master step:}

\State \quad Global consensus update:
\State \qquad 
$m^{k+1} = \left(\alpha N_s I_N + \beta L^{\top} L\right)^{-1} \left(\sum_{i=1}^{N_s} \alpha (p_i^{k+1}+ q_i^k)+\beta L^{\top}(u^k + v^k)\right)$
\State \qquad $u^{k+1} = \mathrm{prox}_{\frac{\mathcal{R}}{\beta}}
\left(Lm^{k+1} -v^k \right)$

\State \textbf{Dual update (parallel):}
\State \quad $q_i^{k+1} = q_i^k+ p_i^{k+1} - m^{k+1}$
\State \quad $v^{k+1} = v^k+ u^{k+1} - Lm^{k+1}$
\State \quad $\lambda_i^{k+1} = \lambda_i^k+ G_i(m^{k+1})-\dobs$
\EndWhile

\State \textbf{Output:} Final model $m$
\end{algorithmic}
\end{algorithm}


\section{Bayesian Inversion Methodology}\label{sec:bayes}
While the deterministic consensus framework established in the previous section provides a computationally efficient path for reconstructing subsurface velocity fields under non-smooth priors, a single point estimate remains fundamentally limited in its ability to characterize the inherent ill-posedness of the tomographic inverse problem. Relying solely on the maximum a posteriori (MAP) or best-fit model ignores the non-uniqueness of the solution and can lead to overconfident interpretations, particularly in regions of the model space with poor illumination or sparse ray coverage. To overcome these limitations, this section extends our distributed architecture into a fully Bayesian framework. By integrating SVGD with the dual-space ADMM mechanism, we move from a point-estimation approach to a particle-based sampling scheme. This hybrid formulation allows leveraging the numerical stability of the AL for uncertainty quantification in high-dimensional model spaces.
\subsection{Bayesian Objective Function}
\medskip
\subsubsection{Likelihood}\label{sec:likelihood}
\medskip
Casting the inverse problem in a Bayesian framework requires specifying the likelihood---i.e., the probability of observing the data given a candidate model. Assuming independent Gaussian measurement noise with known variance $\sigma^2$, the likelihood for source $i$ can be written as
\begin{equation}\label{eq:like}
\pi(\dobs \mid m) \propto \exp\!\Big(\!-\frac{1}{2\sigma^2}\|G_i(m)-\dobs\|_2^2\Big).
\end{equation}
\subsubsection{Prior}\label{sec:prior}
\medskip
The prior distribution $\pi_{\text{prior}}(m)$ encodes geological expectations about the slowness field. We define the regularization functional such that $\pi_{\text{prior}}(m) \propto \exp\!\big(\!-\mathcal{R}(Lm)\big)$, equivalently $-\log \pi_{\text{prior}}(m) = \mathcal{R}(Lm) + \text{const}$. For smooth priors---e.g., Gaussian random fields with Mat\'ern covariance---the gradient $\nabla\mathcal{R}$ exists and can be included directly in the log-posterior gradient of each particle. 
However, many practically relevant priors (e.g., $\ell_1$ sparsity or TV) are non-smooth and do not admit Lipschitz-continuous gradients. While this poses challenges for standard gradient-based Bayesian inference methods, the proposed ADMM-based framework circumvents this difficulty by introducing auxiliary variables and proximal mappings. This effectively separates the non-smooth prior from the data misfit term and replaces the original non-smooth objective with a sequence of easier subproblems (see \cref{m_subprob}), enabling efficient inference even with non-differentiable regularization.
\subsubsection{Posterior}\label{sec:posterior}
\medskip
By Bayes' rule, the negative log-posterior takes the form

\begin{equation}\label{eq:neglogpost}
-\log \pi(m|\tau^\text{obs}) = \sum_{i=1}^{N_s}\frac{1}{2\sigma^2}\|G_i(m)-\dobs\|_2^2 + \mathcal{R}(Lm) + \text{const}.
\end{equation}

In the above expression, the first term plays the role of the negative log-likelihood, and the second corresponds to the negative log-prior. Sampling from this posterior is the central computational challenge, particularly when $\mathcal{R}$ is non-smooth.

\subsection{Stein Variational Gradient Descent}\label{sec:svgd}
\Cref{m_subprob} defines an evolving unnormalized density over ${m}$
 \citep{Siahkoohi_2026_DSP}:
\begin{equation}\label{eq:evolving}
\pi_k(m|\tau^{\text{obs}}) \propto \exp\!\Big(\!-\sum_{i=1}^{N_s}
\frac{\alpha}{2} \|p_i^{k+1} - m + q_i^k\|_{\Sigma^{-1}}^2-
\frac{\beta}{2} \|u^k - Lm + v^k\|_{\Sigma^{-1}}^2\Big),
\end{equation} 
with a proper covariance matrix $\Sigma$.
%

This distribution is entirely smooth in ${m}$, and its gradient can be computed.
Since both terms are smooth and differentiable, SVGD can target $\pi_k$ directly using the above gradient. 

Computing this gradient for each particle reuses the consensus worker step of the deterministic formulation: the local update $p_i^{(j)}$ solves a data-space Gauss--Newton system in which the penalty ratio $\alpha/\mu$ is replaced by its Bayesian-calibrated counterpart $\gamma_i^{(j)}:= \alpha\sigma^2/\mu$, with the adaptive per-source choice 
also available. The resulting scheme is summarized in \Cref{alg:main}.

Unlike the sampling-based methods like MCMC, the SVGD evolves an ensemble of $N_p$ particles
$\{m^{(j)}\}_{j=1}^{N_p}$ to approximate the target posterior distribution. 
The particles are updated iteratively according to:
\begin{equation}
    m^{(j)} \leftarrow m^{(j)} + \eta \phi(m^{(j)}),
\end{equation}
where $\eta$ is the step size, and the optimal perturbation direction is:
\begin{equation}\label{eq:svgd_phi}
    \phi(m^{(j)})=\frac{1}{N_p}\sum_{l=1}^{N_p} {\Big[} K(m^{(l)},m^{(j)})\nabla{_{m^{(l)}}} \log {\pi_k}(m^{(l)}{\mid}\tau^{\text{obs}}) + \nabla{_{m^{(l)}}} K(m^{(l)},m^{(j)}) {\Big]},
\end{equation}
with 
\begin{equation*}
    K(m,m')=\exp\left( -\frac{\|m-m'\|_2^2}{2h^2}\right)
    \Rightarrow
    \nabla_m K = - \frac{1}{h^2}(m-m') K(m,m').
\end{equation*}
The bandwidth is set at every iteration by the median heuristic of \cite{Liu_2016_SVGD}, $h^2=\text{med}^2/\log N_p$, where $\text{med}$ is the median pairwise distance in the current ensemble.
\subsection{Algorithm}\label{sec:algorithm}

Each particle $j$ maintains: model ${m}^{(j)}$, auxiliary variable ${u}^{(j)}$, regularization dual ${v}^{(j)}$, consensus duals ${q}_i^{(j)}$, data multipliers $\Bs{\lambda}_i^{(j)}$, and local copies ${p}_i^{(j)}$. The complete algorithm is summarized in \Cref{alg:main}.

\begin{algorithm}[H]
\caption{ADMM-SVGD for Bayesian traveltime tomography}\label{alg:main}
\begin{algorithmic}[1]
\small
\State \textbf{Input:} Data $\{\Bs{\tau}_i^{\text{obs}}\}_{i=1}^{N_s}$, noise variance $\sigma^2$, prior $\mathcal{R}$, weights $\alpha,\beta$, particles $N_p$, step size $\eta$
\State \textbf{Init:} ${m}^{(j)}\!\sim\! \pi_{0}$, ${u}^{(j)}\!=0$, ${v}^{(j)}\!=\!0$, $\Bs{\lambda}_i^{(j)}\!=\!{0}$, ${q}_i^{(j)}\!=\!{0}$ for all $j,i$
\While{not converged}
\For{$j=1,\dots,N_p$; $i=1,\dots,N_s$} \label{sub_m1} \Comment{\textit{Worker step --- parallel over particles $\times$ sources}}
    \State Solve eikonal (FMM) in ${m}^{(j)}$ $\to$ traveltimes $G_i({m}^{(j)})$ and Jacobian $\B{J}_i^{(j)}$
    \State Compute the penalty parameter $\gamma_i^{(j)}$ 
    \State ${g}_i^{(j)} \leftarrow (\B{J}_i^{(j)})^{\top}\big(G_i({m}^{(j)})-\Bs{\tau}_i^{\text{obs}}+\Bs{\lambda}_i^{(j)}\big) + \gamma_i^{(j)}{q}_i^{(j)}$
    \State $\Delta{p}_i^{(j)} \leftarrow -\frac{1}{\gamma_i^{(j)}}{g}_i^{(j)} + \frac{1}{\gamma_i^{(j)}}(\B{J}_i^{(j)})^{\top}\big(\B{J}_i^{(j)}(\B{J}_i^{(j)})^{\top}+\gamma_i^{(j)}\B{I}_{N_r^i}\big)^{-1}\B{J}_i^{(j)}{g}_i^{(j)}$
    \State ${p}_i^{(j)} \leftarrow {m}^{(j)} + \Delta{p}_i^{(j)}$
\EndFor \label{sub_m2}
\For{$j=1,\dots,N_p$} \label{sub_g1} \Comment{\textit{Gradient assembly (gradient of \cref{eq:evolving}), using updated ${u}$}}
    \State ${z}^{(j)} \leftarrow \frac{1}{N_s}\sum_i{p}_i^{(j)}+\frac{1}{N_s}\sum_i{q}_i^{(j)}$
    \State ${g}_m^{(j)} \leftarrow {\left(\alpha I_N + \beta L^{\top}L\right)^{-1}} \left( \alpha N_s\,({z}^{(j)} - {m}^{(j)}) + \beta\,{L}^{\top}({u}^{(j)}-{L}{m}^{(j)}+{v}^{(j)}) \right)$
\EndFor \label{sub_g2}
\State \label{sub_SVGD1} Compute RBF kernel bandwidth $h$ via median heuristic over $\{{m}^{(j)}\}$
\For{$j=1,\dots,N_p$} \Comment{\textit{SVGD perturbation --- computed over \emph{current} particles}}
    \State $\Bs{\phi}^{(j)} \leftarrow \frac{1}{N_p}\sum_{l}\big[K({m}^{(l)},{m}^{(j)}){g}_m^{(l)} + \nabla_{{m}^{(l)}}K({m}^{(l)},{m}^{(j)})\big]$
\EndFor
\For{$j=1,\dots,N_p$} \Comment{\textit{Particle update}}
    \State ${m}^{(j)} \leftarrow {m}^{(j)} + \eta\,\Bs{\phi}^{(j)}$
    \Comment{\textit{$u$-update---proximal, given current $m$}}
 \State $u^{(j)}\leftarrow  \mathrm{prox}_{\mathcal{R}/{\beta}} \label{sub_u}
\left(Lm^{(j)} -v^{(j)} \right)$
\EndFor \label{sub_SVGD2}
\For{$j=1,\dots,N_p$; $i=1,\dots,N_s$} \label{sub_Dual1} \Comment{\textit{Dual updates}}
    \State ${v}^{(j)} \leftarrow {v}^{(j)} + {u}^{(j)}-{L}{m}^{(j)}$
    \State ${q}_i^{(j)} \leftarrow {p}_i^{(j)}-{m}^{(j)}+{q}_i^{(j)}$
    \State $\Bs{\lambda}_i^{(j)} \leftarrow G_i({m}^{(j)})-\Bs{\tau}_i^{\text{obs}}+\Bs{\lambda}_i^{(j)}$
\EndFor \label{sub_Dual2} 
\EndWhile
\State \textbf{Output:} Posterior samples $\{{m}^{(j)}\}_{j=1}^{N_p}$ approximating $\pi({m}|\Bs{\tau}^{\text{obs}})$
\end{algorithmic}
\end{algorithm}

Every step in \Cref{alg:main} follows directly from the consensus AL of \Cref{sec_deterministicInv}. The ordering is prescribed by ADMM ~\citep{Boyd_2011_DOS}: primal blocks are updated sequentially, each using the most recent values of all other variables. 
\begin{enumerate}
    \item The ${m}$-update minimizes $\mathcal{L}$ over ${m}$ for fixed ($u^{k},v^k,\Bs{\lambda}^k)$: the consensus workers compute the data gradient via the SMW matrix identity in the $N_r^i\times N_r^i$ data space (lines~\ref{sub_m1}--\ref{sub_m2}).
    \item The full gradient of $\log \pi_k$ from \cref{eq:evolving} is assembled using the updated $\u^{k+1}$ (lines~\ref{sub_g1}--\ref{sub_g2}).
    \item SVGD replaces the deterministic minimization with a particle transport step that targets the smooth distribution $\pi_k$ (lines~\ref{sub_SVGD1}--\ref{sub_SVGD2}).
    \item The $u$-update (line~\ref{sub_u}) minimizes $\mathcal{L}$ over $u$ for fixed $({m}^k,v^k)$ via the proximal operator.
    \item The dual updates then perform the standard dual ascent (lines~\ref{sub_Dual1}--\ref{sub_Dual2}).
\end{enumerate}
The computational cost per iteration is $O(N_p N_s)$ parallel eikonal solves.\\
\medskip
\textbf{Inner and outer loops.} In practice, \Cref{alg:main} is structured as a two-level iteration. The \emph{outer loop} updates the Lagrange multipliers $\Bs{\lambda}_i$ and $v$, which define the evolving target distribution $\pi_k$ in \cref{eq:evolving}. The \emph{inner loop} performs multiple SVGD steps (together with $u$-updates) for a fixed set of multipliers, approximately transporting the particles toward $\pi_k$ before the target changes. Heuristically, running more inner SVGD steps allows the particle ensemble to better approximate the current evolving posterior before the multiplier updates shift the target to $\pi_{k+1}$. This is consistent with the inexact ADMM framework~\cite{Boyd_2011_DOS}, which permits approximate primal solves provided the approximation improves over iterations---a condition naturally satisfied by SVGD through warm-starting, since the particles from the previous outer iteration already provide a good initialization for the updated target.

\graphicspath{{./figs/}}

\section{Numerical examples}\label{sec:results}
\medskip
We assess the proposed framework on three test problems of increasing dimension, summarized in \Cref{tab:examples}. (I) The one-dimensional (1D) vertical seismic profile (VSP) (\Cref{sec:vsp1d}), small enough that a converged MCMC posterior remains affordable and can serve as the reference against which the ADMM-SVGD ensemble is calibrated.
(II) The two-dimensional (2D) 2004 BP salt benchmark (\Cref{sec:bp2004}) is run noise-free against the deterministic consensus ADMM of \Cref{alg:det}, and then under three noise levels, to assess accuracy and noise robustness. (III) The three-dimensional (3D) SEG/EAGE salt model (\Cref{sec:seg3d}), an order of magnitude larger. The settings shared by all examples are collected in \Cref{sec:num_common}; only the model-specific settings are given in the individual sections.    
\subsection{Common settings}\label{sec:num_common}
\medskip
In every experiment the eikonal equation~\eqref{eikonal} is solved with the same second-order FMM of \Cref{sec:FMM}, on a uniform grid shared by the forward solver and the inversion. 
The Jacobian $J_i^k$ is assembled by backtracking the geodesic rays from each receiver along $-\nabla T$. 
Observed traveltimes are generated with the same solver in the true model, 
$\tau^\text{obs}=
\begin{bmatrix}
    G_1 (m^\text{true})& G_2 (m^\text{true})&\cdots & G_{N_s} (m^\text{true})
\end{bmatrix}
$, and, where indicated, contaminated with additive Gaussian noise
\begin{equation}\label{eq:noise_model}
    \tau^\text{obs} = \tau^\text{clean}+\varepsilon, \qquad \varepsilon \sim \mathcal{N} \! \left(0, \, \sigma_\text{noise}^2  \right), \qquad \sigma_\text{noise} = \nu \, \bar{\tau}^\text{clean},
 \end{equation}
where $\bar{\tau}^{clean}$ is the mean clean traveltime over all source-receiver pairs and $\nu$ is the noise fraction. 
Solver, grid, and data are identical for the deterministic and Bayesian runs. Therefore, any difference between them is attributable to the inversion scheme alone. We invert for slowness $m=1/v$, with $v$ in km/s, and convert back to velocity before display. TV regularization is used in all experiments. 

\paragraph{SVGD step size.}
The Stein direction $\phi^{(j)}$ of \eqref{eq:svgd_phi} is applied with a step normalized to the length of the ADMM likelihood gradient $g_\text{like}^{(j)}$, 
\begin{equation}\label{eq:svgd_step}
    m^{(j)} \leftarrow m^{(j)} + \eta^{(j)} \, \phi^{(j)}, \qquad \eta^{(j)} = \xi \, \frac{\|g_\text{like}^{(j)} \|_2}{\|\phi^{(j)}\|_2}, \qquad \xi \leq 1,  
\end{equation}
so that the method retains the scaling of the ADMM update, which is set by $\mu$ through~\eqref{eq:smw}, while replacing its direction by the SVGD one. The value of $\xi$ is given for each experiment in the corresponding section. 

\paragraph{Initialization.}
The deterministic runs start from the fitted 1D gradient model $v_0=a+bz$ of \Cref{sec:init} and the particle ensembles are drawn around $v_0$, from a fixed seed using a Gaussian random field. All experiments were carried out on a dual Intel Xeon Platinum 8176 system, 56 cores at 2.10 GHz. The $N_p\times N_s$ local subproblems of each outer iteration are distributed across 30 workers using a parallel processing pool. 

\paragraph{Quality measures.} The following diagnostics are monitored throughout. The data fit of particle $j$ at iteration $k$ and the accuracy of the posterior mean are measured by the relative data residual and the relative model error (RME)
\begin{equation}
 \mathcal{E}_d^{(j)}(k) = \frac{\| \tau^\text{obs}-G(m^{(j),\, k})  \|_F}{ \| \tau^\text{obs} \|_F}, 
\qquad 
\text{RME}(k) = 100 \times \frac{ \|v^\text{true}-\bar{v}^{\, k} \|_2}{\|v^\text{true}\|_2} \ [\%],
\end{equation}
where
$G(\cdot)=
\begin{bmatrix}
    G_1 (\cdot)& G_2 (\cdot)&\cdots & G_{N_s} (\cdot)
\end{bmatrix} \in \mathbb{R}^{N_r\times N_s}
$
collects the traveltimes for all of the $N_s$ sources, with $\|\cdot\|_F$ denoting the Frobenius norm,  and $\bar{v}^{\, k} = \frac{1}{N_p}\sum_{j}v^{(j),\, k}$ is the ensemble mean, which reduces to the single deterministic model for $N_p=1$; both are evaluated in the velocity domain. 
Progress of the AL loop towards feasibility is tracked through $\|\lambda^{(j),\, k}\|_F$, which plateaus once the traveltime residuals that increment it become negligible. Uncertainty is reported as the pointwise ensemble standard deviation together with the 95\% credible interval of the final ensemble at selected profiles. 
In the noise study of \Cref{sec:bp_noise} we additionally report the inter-realization standard deviation of the posterior mean, i.e., the pixelwise spread of the posterior means obtained from independent noise draws. 
In all experiments below the iteration budget is fixed at $K$ outer iterations rather than set by a convergence test, and the reported ensemble should be understood as the state at the budget rather than the converged one. 
\begin{table}[t]
\centering
\caption{Overview of the three numerical experiments. $N$ is the number of
inverted model parameters, $N_s$ and $N_r$ the number of sources and
receivers, and $N_p$ the number of particles ($N_p=1$ denotes the
deterministic baseline of \Cref{alg:det}).}
\label{tab:examples}
\small
\begin{tabular}{@{}llccccl@{}}
\toprule
Section & Model & $N$ & $N_s$ & $N_r$ & $N_p$ & Purpose \\
\midrule
\ref{sec:vsp1d} & 1D VSP                & $41$      & $1$   & $20$  & $1000$  & Calibration against MCMC \\
\ref{sec:bp2004}& 2D BP 2004 salt       & $36\,000$ & $67$  & $225$ & $100$     & Accuracy, noise robustness \\
\ref{sec:seg3d} & 3D SEG/EAGE salt      & $434\,146$& $100$ & $154$ & $100$      & Scalability, 3D feasibility \\
\bottomrule
\end{tabular}
\end{table}

\begin{figure}
\centering
\includegraphics[width=0.5\textwidth]{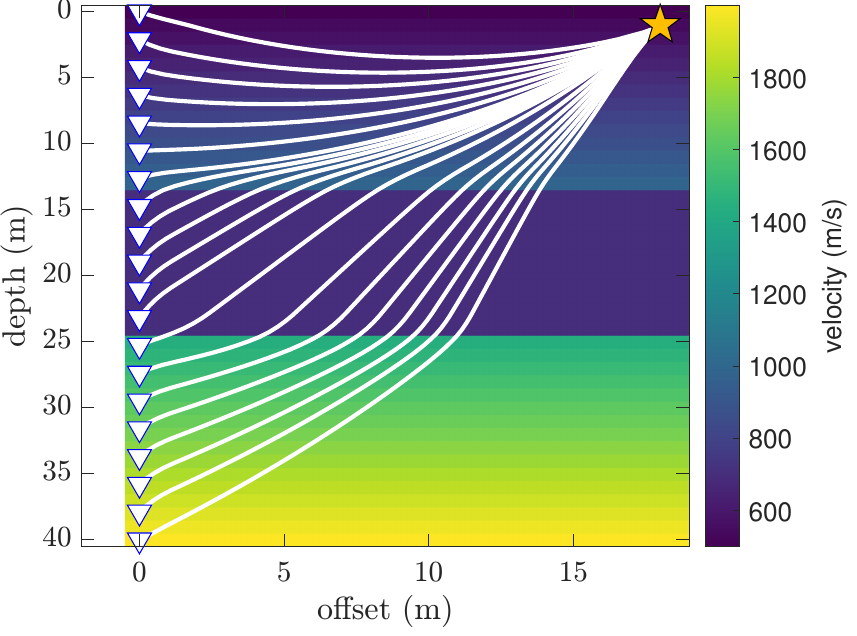}\hspace{0.035\textwidth}
\caption{1D VSP test: acquisition and ground truth, with the source (star) and the
receiver well (triangles).}
\label{fig:setup}
\end{figure}

\begin{figure}
\centering
\includegraphics[width=0.82\textwidth]{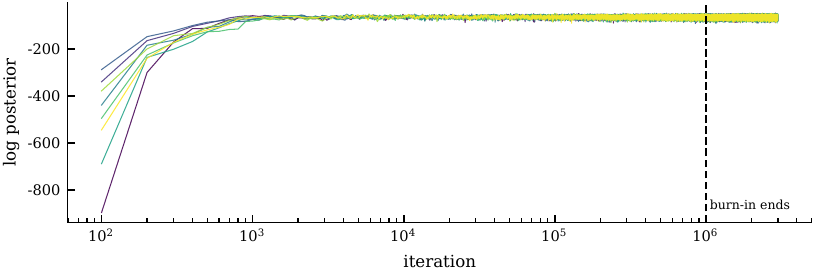}
\caption{Convergence of MCMC reference chains. Log-posterior against iteration through burn-in, each
chain in its own color, with the end of burn-in  marked by a dashed line. Every chain reaches the typical set
within the first few thousand iterations, and the retained draws give a rank-normalized
split-$\hat{R}$ of $1.0002$ across all coordinates.}
\label{fig:mixing}
\end{figure}

\begin{figure}
\centering
\includegraphics[width=0.88\textwidth]{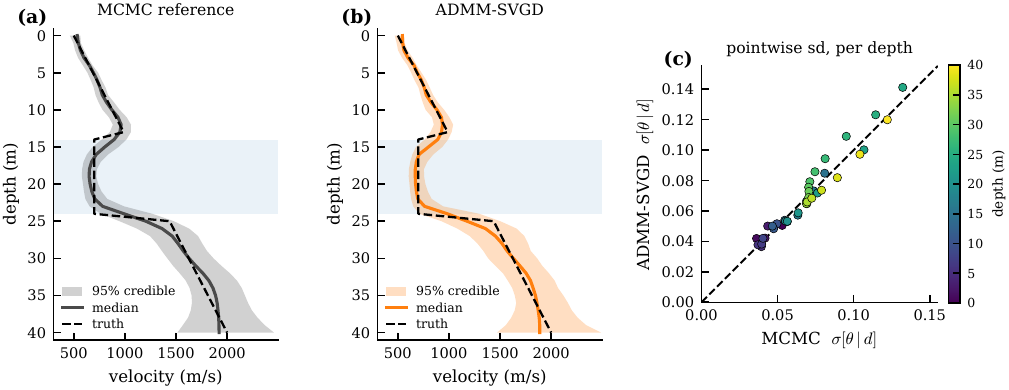}
\caption{1D VSP test: ADMM-SVGD against the MCMC reference. (a) Reference posterior: median velocity
inside its $95\%$ credible band, against ground truth, with the low-velocity layer shaded.
(b) The ADMM-SVGD ensemble, on the same axes. (c) Pointwise posterior standard deviation of
the ensemble against that of the reference, one point per depth.}
\label{fig:results}
\end{figure}
\subsection{1D vertical seismic profile (VSP) problem}\label{sec:vsp1d}
\medskip
First, we consider a small nonlinear 1D example as an affordable calibration test of the paper, in which we assess the consensus ADMM-SVGD against a reference MCMC that can be trusted. 
\subsubsection{Experimental design}
\medskip
The true model is a 1D profile, in which the velocity rises linearly from $500$ to $2000$~m/s over $40$~m, with $700$~m/s low-velocity layer between $14$
and $24$~m depth (\cref{fig:setup}). A single source sits near the surface at ($18,\,1$)~m and $N_r=20$~receivers are distributed down a well at $x=0$~m. The unknown profile $m \in \mathbb{R}^{41}$ is replicated laterally across the $41 \times 20$ grid on which the traveltime field is computed. The ensemble uses $N_p=1000$ particles, and inverts for the unknown on a grid four times coarser than the one on which the data are generated. 

Two properties make the example informative despite its size. The Jacobian has rank at most $20$ against
$41$ parameters, so roughly half of the model space is prior-determined, and a sampler
miscalibrated in the null space cannot hide behind a good data fit. Illumination also decays
monotonically with depth, since a single near-surface source illuminates receivers below it,
so we expect the posterior to be widened toward the base of the well.
\subsubsection{Results}
We sample \eqref{eq:neglogpost} with preconditioned adaptive Metropolis \cite{Haario_2001_AAM}, in which the proposal shape is the Gauss-Newton form, and adaptation is confined to burn-in, so that the retained draws are time-homogeneous. Eight chains start over-dispersed on the posterior scale, five standard deviations from the deterministic solution, and as the \cref{fig:mixing} shows, every chain reduces the typical set well within the burn-in.
The retained draws clear the modern rank-normalized  split-$\hat{R}$ and effective sample size thresholds on every coordinate with room to spare \cite{Vehtari_2021_RNF}. Therefore, the comparison below is a statement about the sampler rather than about the baseline. 

\Cref{fig:results} places the two posteriors side by side. The median profiles (panels a and b) are
difficult to distinguish, and both place the ground truth inside their credible band over
essentially the whole section. The agreement extends to the second moment:
the pointwise standard deviations agree in level and, more tellingly, in
depth structure, widening toward the base of the well as illumination falls (panel c). Recovering this
structure, rather than a uniform width, is the substantive observation.
That the spread is not an artifact of the initialization was checked by repeating the run from an under-dispersed ensemble and from one at the prior scale (not shown here); both converge to a common level.

\subsection{2D problem: BP 2004 model}\label{sec:bp2004}
We now turn to a 2D benchmark, the 2004 BP salt velocity model \citep{billette20052004} introduced in \cref{fig:BP_rays}, whose combination of smoothly varying sediments with strong, high-contrast salt bodies stresses both the nonlinearity of the forward map and the ability of the prior to preserve sharp interfaces.
The free parameters in model subproblem \eqref{m_subprob} are fixed at $\alpha=1$ and $\beta=5$. Furthermore, in the SVGD step~\eqref{eq:svgd_step} we use $\xi=0.2$ for the first ten iterations and $\xi=0.05$ thereafter. 
Two experiments are reported: a noise-free run (\Cref{sec:bp_noisefree}) that compares ADMM-SVGD (\Cref{alg:main}) against the deterministic consensus ADMM (\Cref{alg:det}), and a repetition of the inversion at three noise levels with ten realizations each (\Cref{sec:bp_noise}). Both share the setup described next.  
%
%
\subsubsection{Experimental design}\label{sec:bp_setup}
\begin{figure}
    \centering
    \includegraphics[width=0.7\linewidth]{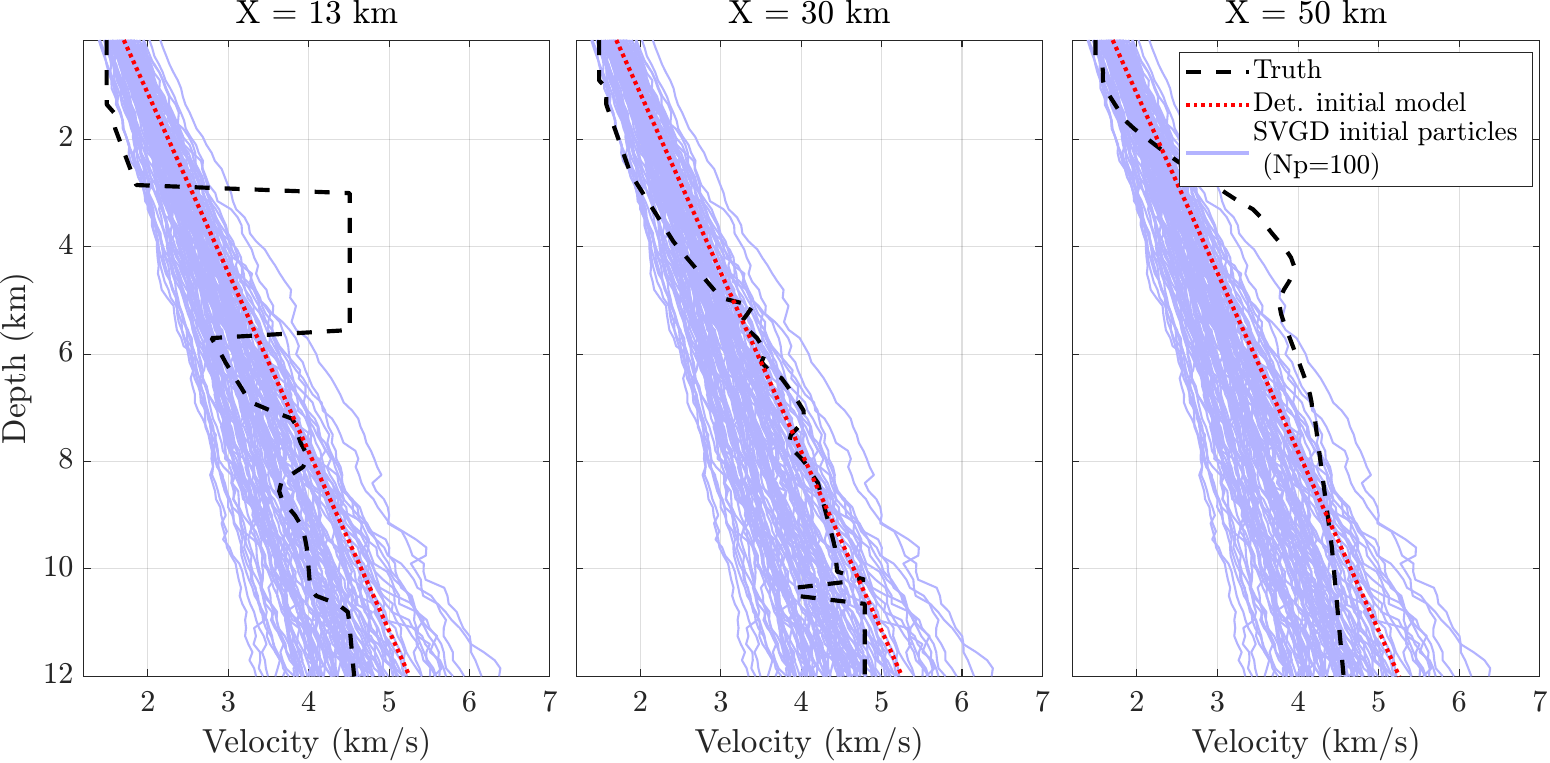}
    \caption{BP 2004 model: the initial particle ensemble. Vertical velocity profiles at $X=13,~30,~50$~km showing the true model, the deterministic initial model, and the 100 SVGD initial particles. }
    \label{fig:BP_initials}
\end{figure}

\medskip
\paragraph{Discretization  and acquisition.}
The inversion and the forward solves share a single uniform grid of $N_z \times N_x = 80 \times 450$ nodes with $\Delta x=\Delta z = 150$~m over the $12 \times 67.5$~km domain. 
The acquisition is an ultra-long-offset, fixed-spread OBS geometry: $N_s=67$ sources along the sea surface between $x=0.15$~km and $x=67.2$~km, and a fixed spread of $N_r=225$ seafloor receivers at $0.3$~km interval spanning 
$x=0.1-67.35$~km. Every receiver records every source,
so the full data set consists of $N_r\times N_s$ first arrival traveltimes and each source contributes a data space Gauss-Newton system of size only $225 \times 225$, to be compared with the $N\times N$ (with $N=N_z\times N_x$) model-space Hessian that the SMW identity~\eqref{eq:smw} allows us to avoid. The geometry is deliberately one-sided: sources and receivers are confined to the top of the model, so deeper parts of the model, edges, and large portions of the sub-salt region are weakly sampled or unsampled, as quantified by the illumination map of \cref{fig:BP_rays} (bottom). \\  
For the SVGD runs, an ensemble of $N_p=100$ particles $\{ m^{(j)} \}_{j=1}^{N_p}$ is drawn. \Cref{fig:BP_initials} shows vertical profiles of the resulting ensemble against the true model and the deterministic initial model. 
\begin{figure}
    \centering
    \includegraphics[width=0.7\linewidth]{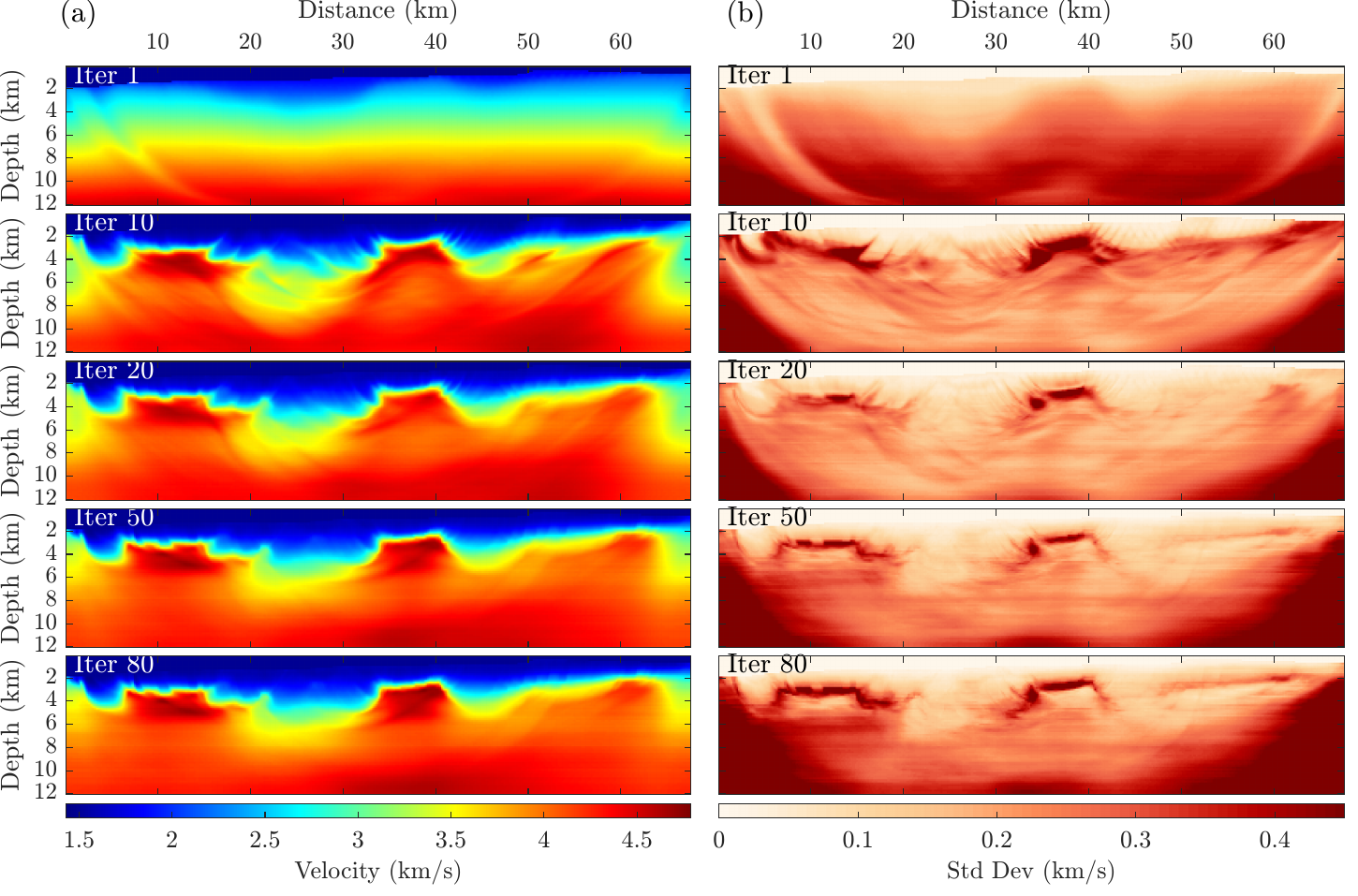}
    \caption{BP 2004 model (noise-free). ADMM-SVGD: Evolution of the posterior estimates of the velocity model with iteration number. (a) Posterior mean velocity models. (b) Corresponding posterior standard deviation models.}
    \label{fig:BP_svgd_mean_std_vs_iter}
\end{figure}

\begin{figure}
    \centering
    \includegraphics[width=0.9\linewidth]{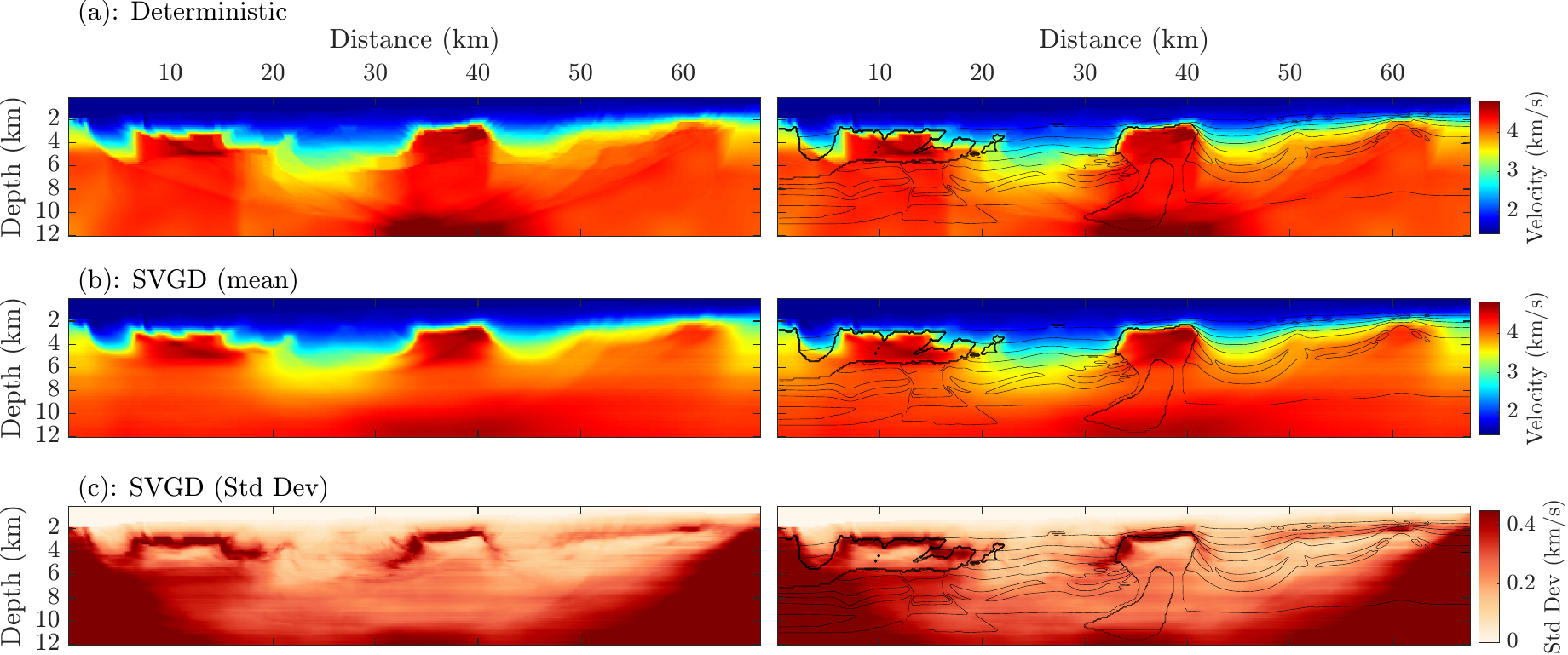}
    \caption{BP 2004 model (noise-free, final iteration). Comparison of the deterministic ADMM and ADMM-SVGD results, shown without (left column) and with (right column) the true model contours overlaid. (a) Deterministic ADMM velocity model. (b-c) ADMM-SVGD with (b) posterior mean velocity model and (c) posterior standard deviation model.}
    \label{fig:BP_det_vs_svgd_mean_std}
\end{figure}
\begin{figure}
    \centering
    \includegraphics[width=0.7\linewidth]{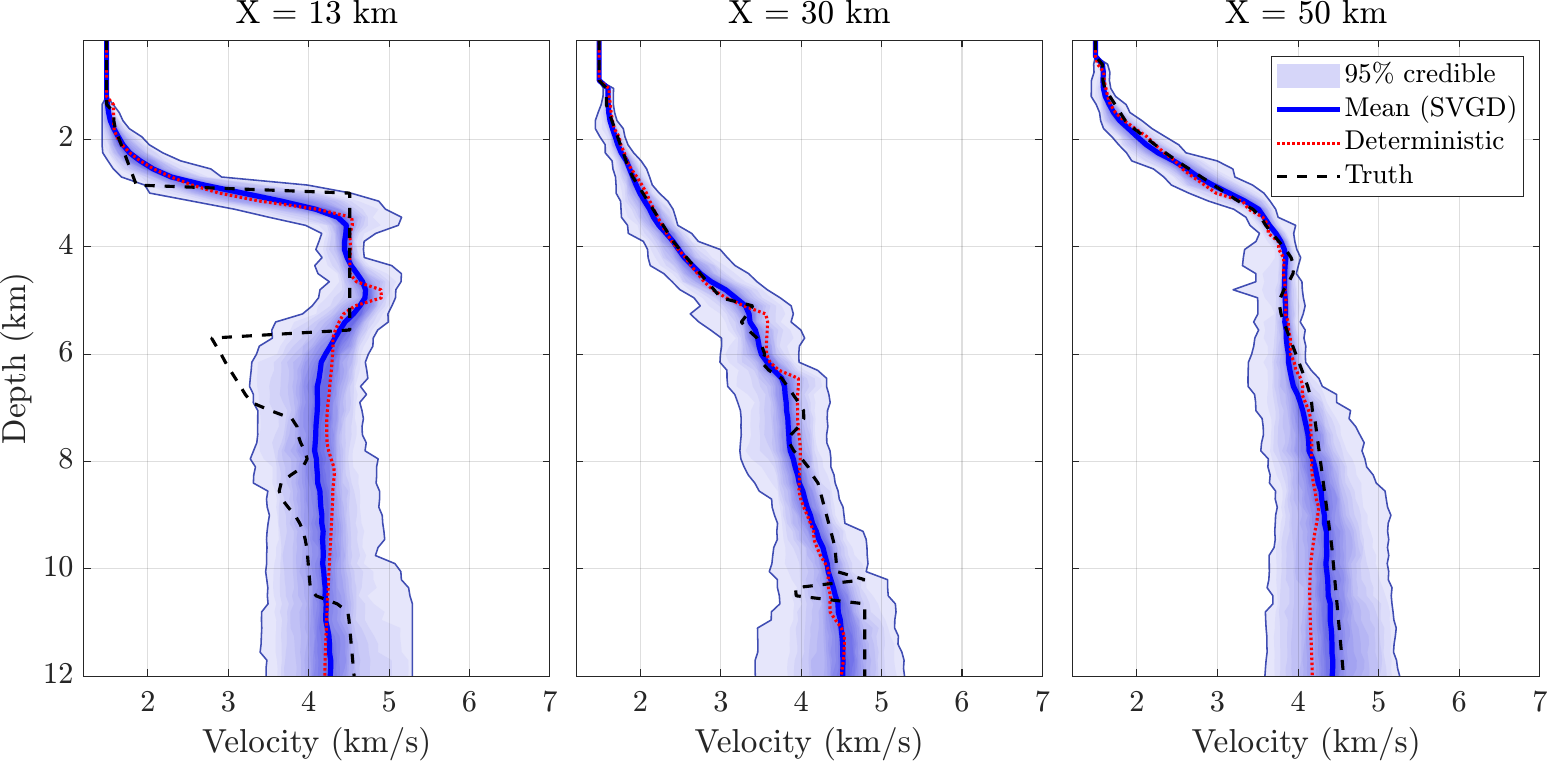}
    \caption{BP 2004 model (noise-free). Vertical velocity profiles at $X=13,~30,~50$~km comparing the true model, deterministic result, and SVGD posterior mean with its 95\% credible interval, corresponding to the results shown in \cref{fig:BP_det_vs_svgd_mean_std}. Within the band, the shading encodes the density of the particle ensemble at each depth: darker shading marks the regions of high posterior density, and shading fades toward the edges of the 95\% interval.}
    \label{fig:BP_svgd_CI}
\end{figure}
\begin{figure}
    \centering
    \includegraphics[width=1\linewidth]{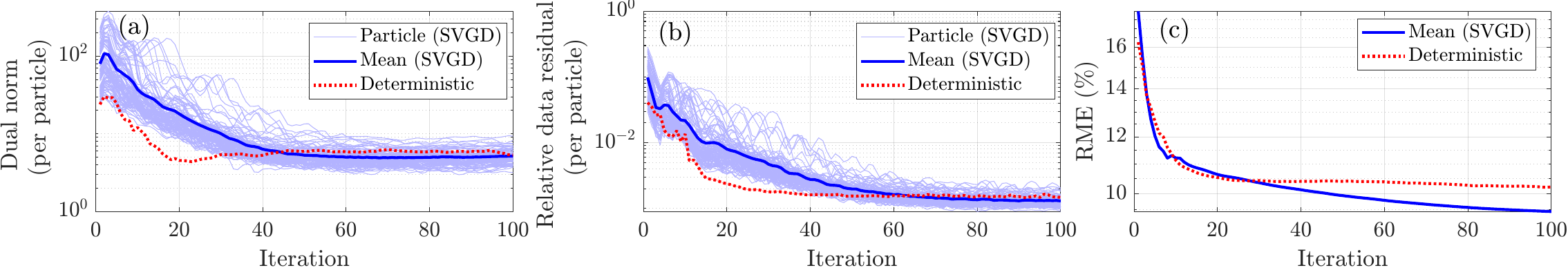}
    \caption{BP 2004 model (noise-free). Convergence diagnostics for ADMM-SVGD compared to the deterministic (single-model) baseline. Thin light-blue lines: individual particles; solid blue: SVGD ensemble mean; dotted red: deterministic. (a) Dual variable norm per particle. (b) Relative data residual per particle. (c) RME of the posterior mean velocity vs. iteration.}
    \label{fig:BP_svgd_errors}
\end{figure}
\subsubsection{Noise-free data: ADMM-SVGD versus deterministic ADMM}\label{sec:bp_noisefree}
\medskip
We first compare the two algorithms on the data generated in the true model, without added noise. \Cref{fig:BP_svgd_mean_std_vs_iter} follows the ensemble mean and the pointwise standard deviation through the iterations. At the first iteration, the mean is still essentially the 1D starting trend beginning to shape the salt bodies, and the standard deviation shows little imprint of the salt: it is a smooth bowl, lowest in the shallow center, and rising towards the lateral edges and the base.  
By iteration \#10, the two salt bodies have emerged, with their top interface close to its final position. The standard deviation changes character at the same time. The smooth bowl breaks up into ridges along the top-of-salt reflectors and the salt flanks, crossed by fine ray-path smearing (most likely the footprint of the source-wise decoupling of the $\{p_i\}$-update). The ridges mark where the particles disagree about the velocity contrast at the interface. After a sufficient multiplier update in later iterations, the ray footprints fade as $\lambda_i^{(j)}$ enforces consistency across sources. The wedges at both lateral edges behave differently. Since no ray samples these cells, the likelihood gradient vanishes and nothing balances the repulsion in \eqref{eq:svgd_phi} except the prior score: the ensemble slowly diffuses outward. \\
The comparison of the deterministic reconstruction with the posterior mean and standard deviation (with and without the true model overlaid) is shown in \cref{fig:BP_det_vs_svgd_mean_std}. Both estimates place the two shallow salt bodies at the correct depth and lateral extent. Two differences stand out. First, the deterministic model still carries ray-path footprints, whereas the posterior mean is largely free of them. Second, the deterministic salt is slightly blockier at the top and slightly more diffuse at the flanks. \\
\Cref{fig:BP_svgd_CI} shows the same result as a function of depth at the profiles shown in \cref{fig:BP_initials}. The posterior mean and the deterministic profile show good recovery of the true model down to about 6~km depth. The 95\% credible band tracks the illumination: it is tight in the shallow sediment, widens across the top of salt, and reaches over a kilometer per second in the deep section. Across the model, the band contains the true velocity, but the coverage is not uniform. \\
\Cref{fig:BP_svgd_errors} shows the three monitored quantities and makes the difference between the two schemes quantitative. The dual norm (panel a) starts an order of magnitude higher for the ensemble, whose particles begin far from feasibility, but both settle on the same plateau, which suggests a comparable level of constraint satisfaction. The data residual (panel b) is lower for the deterministic run over the first sixty iterations; the ensemble mean then overtakes it; and the spread between the worst- and the best-fitting particles contracts fourfold, so the ensemble converges as a whole rather than being carried by a few members. The model error (panel c) separates them: the deterministic curve stalls after about twenty iterations, whereas the ensemble is still improving at the iteration budget.  
%
%
\begin{figure}
    \centering
    \includegraphics[width=0.9\linewidth]{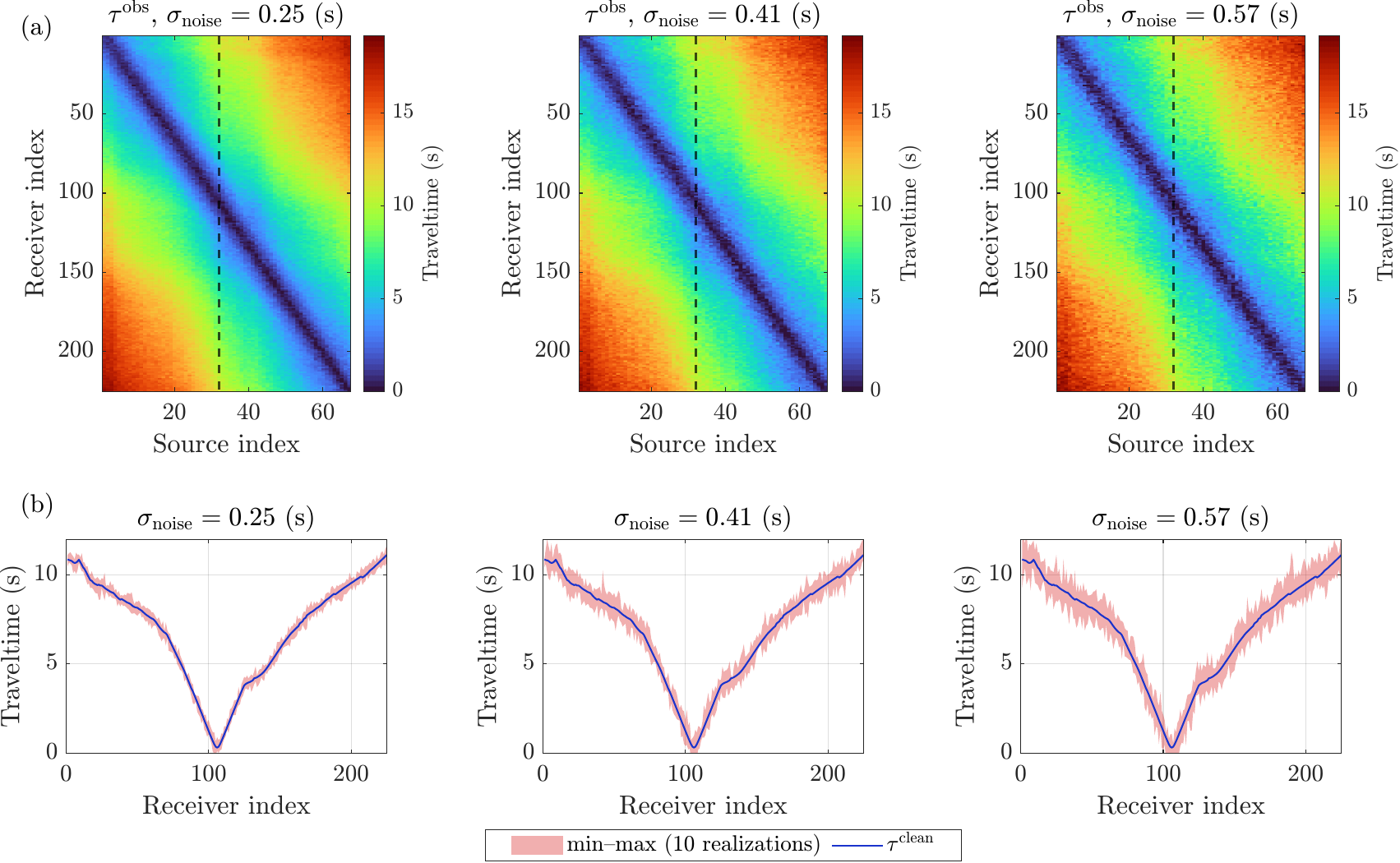}
    \caption{BP 2004 model (noise test). (a) Full noisy traveltime map $\tau^\mathrm{obs}$ (source-receiver coordinates) for each noise level labeled with $\sigma_\mathrm{noise}$; the dashed line marks source \#32, detailed in the row below. (b) Traveltime vs. receiver index for source \#32, showing the pointwise minimum-maximum envelope of $\tau^\mathrm{obs}$ over the 10 independent noise realizations (shaded band), with the clean curve $\tau^\mathrm{clean}$ overlaid. }
    \label{fig:noisy_data}
\end{figure}
\begin{figure}
    \centering
    \includegraphics[width=0.7\linewidth]{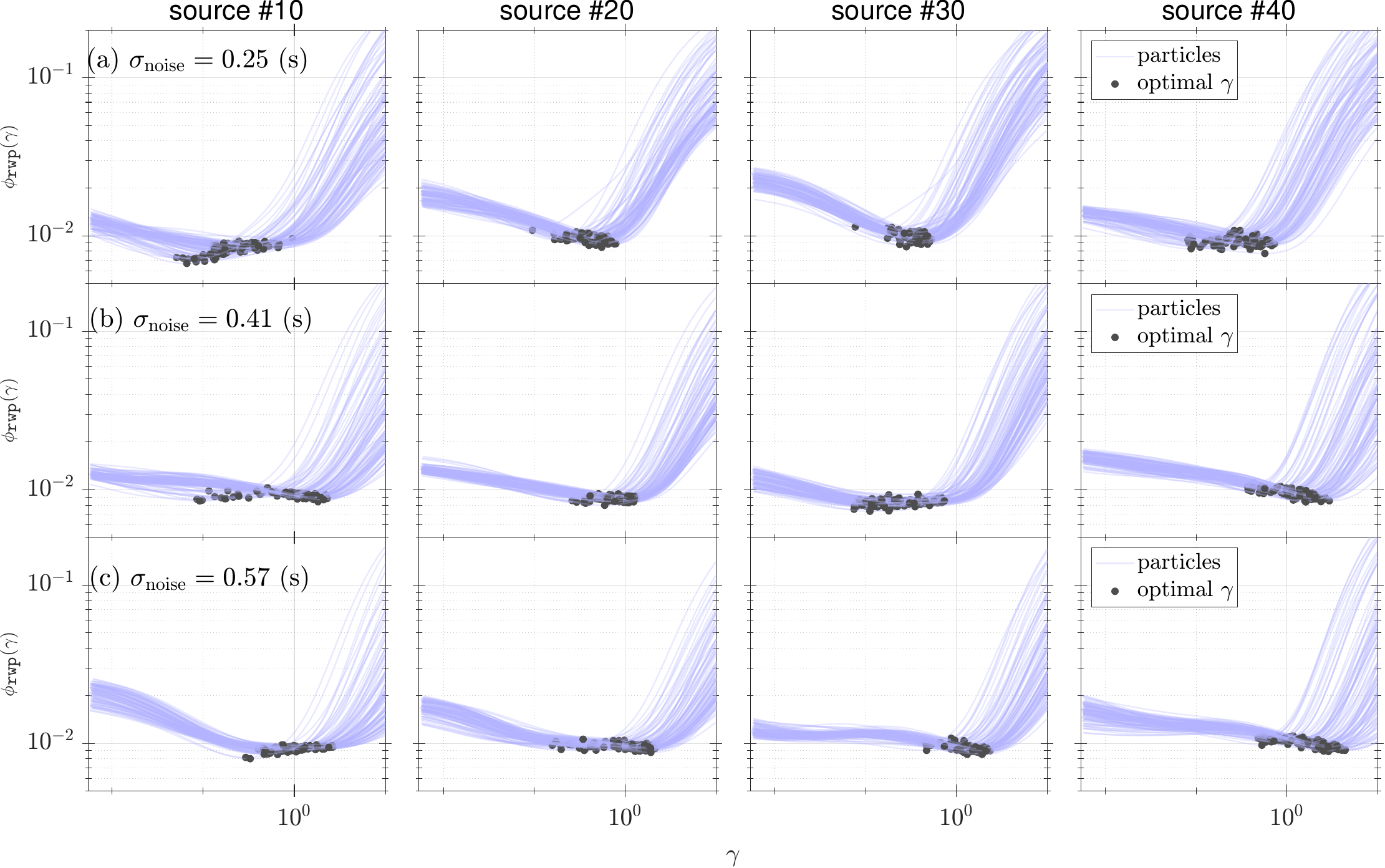}
    \caption{BP 2004 model: data-driven selection of the penalty parameter under noise at the first iteration. Residual-whiteness score $\phi_\texttt{rwp}(\gamma)$ against the penalty parameter $\gamma$ for four representative sources (columns; \#10, \#20, \#30, \#40) across three noise levels (rows; $\sigma_\mathrm{noise}=0.25,~0.41,~0.57$~s). Light-blue curves are per-particle whiteness scores; black dots mark the RWP-selected optimum $\gamma$. The functional $\phi_\texttt{rwp}$ is defined in \Cref{app:B}. }
    \label{fig:BP_gamma_anal}
\end{figure}
\begin{figure}
    \centering
    \includegraphics[width=0.8\linewidth]{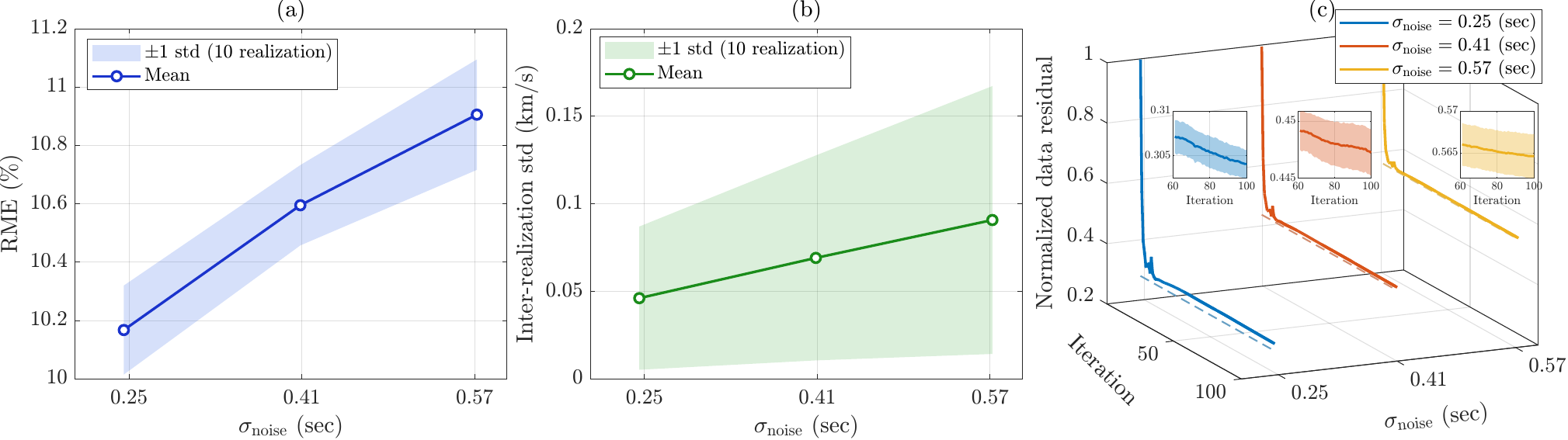}
    \caption{BP 2004 model (noise test). (a) Final RME (\%) vs. $\sigma_\mathrm{noise}$ (mean $\pm 1$ std, 10 realizations). (b) Inter-realization standard deviation of the posterior mean velocity (km/s) vs. $\sigma_\mathrm{noise}$ (mean $\pm 1$ std). (c) Normalized data residual vs. iteration and $\sigma_\mathrm{noise}$. The dashed lines indicate the noise level.}
    \label{fig:Inter_realization_error}
\end{figure}
\begin{figure}
    \centering
    \includegraphics[width=0.6\linewidth]{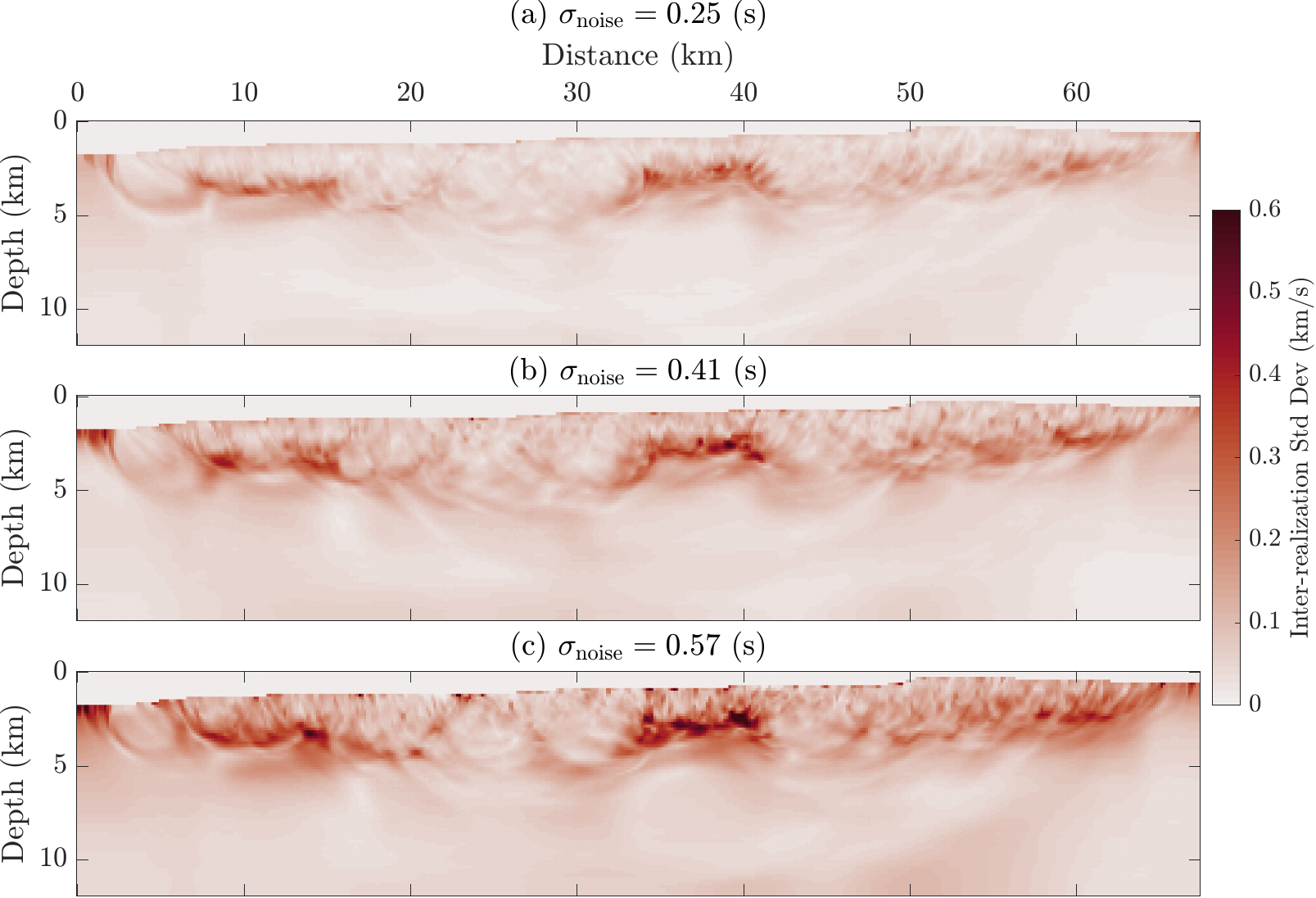}
    \caption{BP 2004 model (noise test). Inter-realization standard deviation of the posterior mean velocity map (km/s), computed pixel-wise across the 10 independent noise realizations at each noise level. This quantifies the inter-realization repeatability of the ADMM-SVGD mean estimate under different noise draws, distinct from the within-run SVGD posterior uncertainty shown in \cref{fig:BP_det_vs_svgd_mean_std}. The scalar, spatially-averaged version of this metric is shown in \cref{fig:Inter_realization_error} (b).}
    \label{fig:Inter_realization_map}
\end{figure}
\subsubsection{ADMM-SVGD: analysis of additive random noise}\label{sec:bp_noise}
\medskip
We now examine how the sampler responds to the noise in the data. The inversion is repeated for three noise levels, $\sigma_\mathrm{noise} \in \{0.25,\, 0.41,\,0.57$\}~s, corresponding to $3\%$, $5\%$ and $7\%$ of the mean clean traveltime. For each level, 10 independent realizations of the noise are generated. Only the ADMM-SVGD algorithm is run here; the settings are those of \Cref{sec:bp_setup} and are held fixed across all runs. In addition, the same stored initial particles are reused throughout, so the differences between runs can be attributed to the noise realization alone. 

\Cref{fig:noisy_data} shows the effect of the noise on the observations. Panel (a) displays the full $N_r \times N_s$ traveltime map at each level. Panel (b) shows a comparison for source~\#32, where the shaded band is the pointwise minimum-maximum envelope of 10 realizations around the clean curve.  
\Cref{fig:BP_gamma_anal} examines whether the data-driven selection of $\gamma_i^{(j)}$ remains well-posed as the noise grows. Each panel shows the residual whiteness score $\phi_\texttt{rwp}(\gamma)$ of  \Cref{app:B} as a function of $\gamma$ for one source, with one light-blue curve per particle, and marks the selected optimum. Two observations follow. First, at every noise level and for each of the four representative sources, the whiteness score has a distinct minimum. Second, the selected $\gamma$ moves systematically to larger values as $\sigma_\mathrm{noise}$ increases, which is the expected and desirable behavior: noisier data are fitted less aggressively, and the Gauss-Newton system is damped more strongly, without any global schedule or manual retuning. 
\Cref{fig:Inter_realization_error}~(a) reports the final model error of the posterior mean as a function of $\sigma_\mathrm{noise}$, averaged over 10 realizations with a $\pm 1$ standard deviation band. The degradation is gradual and approximately linear over the three levels tested: raising the noise from $3\%$ to $7\%$ of the mean clean traveltime costs less than one percentage point of the relative model error, and the whole sweep stays within about 1.5 points of the noise-free results of \cref{sec:bp_noisefree}. 
Panel (b) asks the complementary question: not how far the estimate is from the truth, but how reproducible it is when only the noise draw changes. It plots the inter-realization standard deviation of the posterior mean, that is the pixelwise spread of the 10 recovered mean models, spatially averaged. This quantity is distinct from the within-run posterior standard deviation of \cref{fig:BP_det_vs_svgd_mean_std} (c). The former measures the sensitivity of the estimator to the data; the latter, the width of the posterior recovered by a single run. The inter-realization spread also grows roughly in proportion to $\sigma_\mathrm{noise}$, but its relative increase is much larger than that of the error in panel~(a). The normalized data residual against iteration is shown in panel (c). At every level tested, it drops sharply over the first ten iterations, reaches a plateau close to the dashed noise line at about iteration 20, and stays there. The algorithm therefore fits the data down to the noise floor: the multipliers enforce the traveltime constraints to the accuracy the data support, and the estimated $\gamma_{i}^{(j)}$ appears to prevent the Gauss-Newton step from fitting beyond it. This behavior is consistent with the mild degradation of accuracy in panel~(a) and with the  analysis given in \cite{aghazade2025automatic}.     
\Cref{fig:Inter_realization_map} shows where in the model that inter-realization spread lives. The pattern is the same at all three noise levels, only rescaled. It is concentrated in the upper 5~km, along the top and flanks of the salt and is very low in the deeper part. The noise perturbs the position of the interfaces that the data actually resolve, and leaves the deep, prior-dominated part of the model largely unaffected. 
%
\begin{figure}
    \centering
    \includegraphics[width=0.7\linewidth]{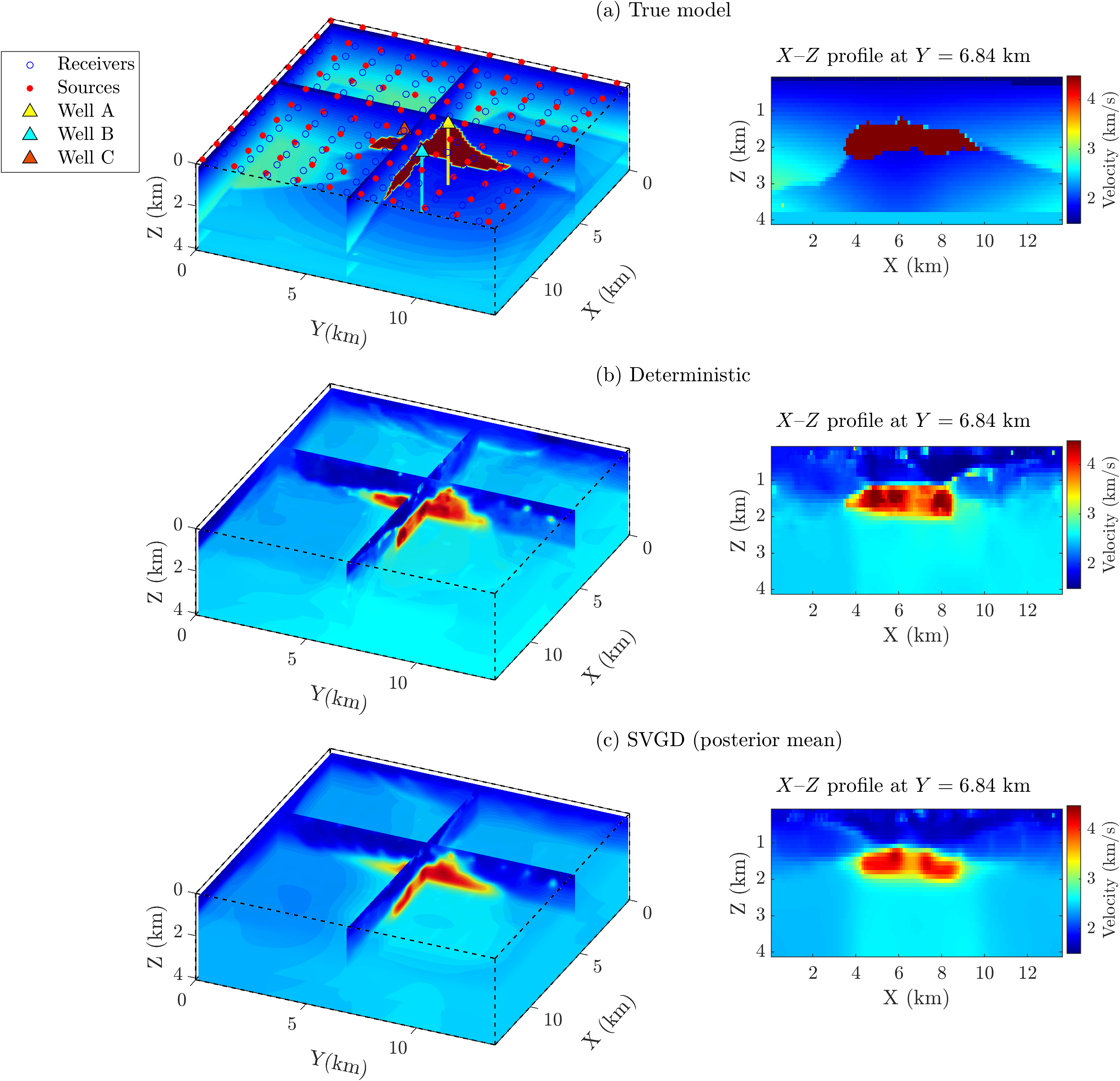}
    \caption{SEG/EAGE 3D salt model: true model and the two reconstructions. Each row shows orthogonal slices through the volume (left) and the $X-Z$ section at $Y=6.78$~km (right). (a) True model, with locations of wells A-C and the surface acquisition geometry overlaid (red stars: sources; blue circles: receivers). (b) Deterministic consensus ADMM reconstruction. (c) ADMM-SVGD posterior mean.}
    \label{fig:3D_true_init_det}
\end{figure}
\begin{figure}
    \centering
    \includegraphics[width=0.7\linewidth]{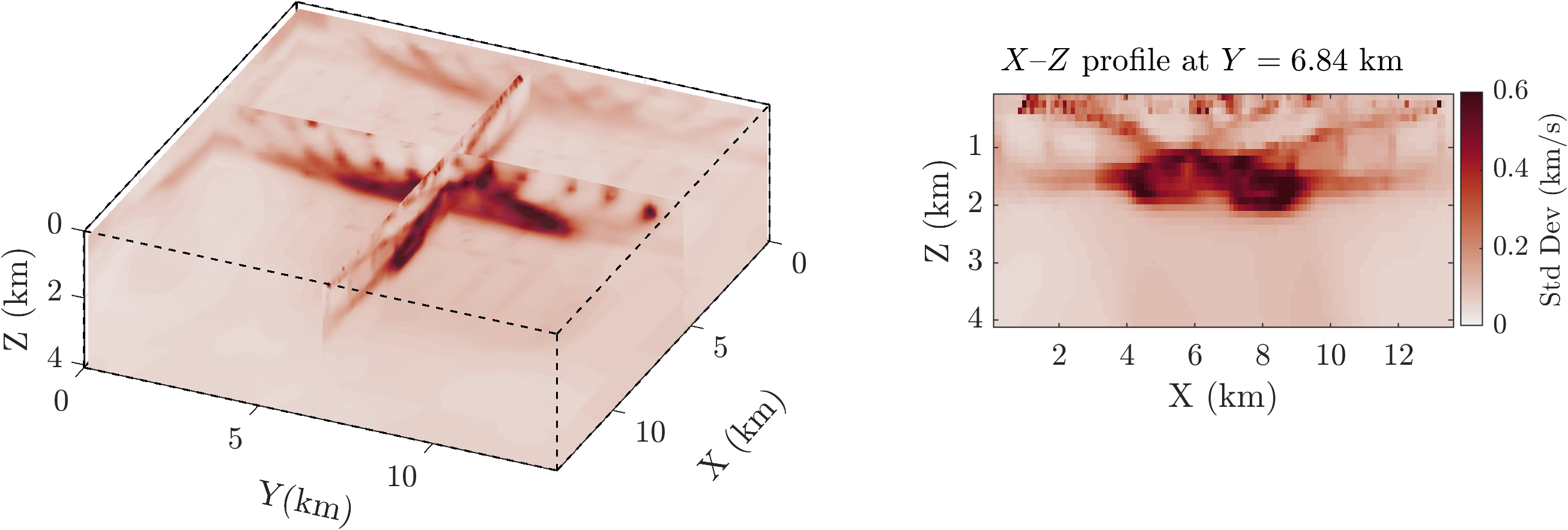}
    \caption{SEG/EAGE 3D salt model: posterior uncertainty. ADMM-SVGD pointwise posterior standard deviation, and its $X-Z$ profile.}
    \label{fig:3D_svgd}
\end{figure}
\begin{figure}
    \centering
    \includegraphics[width=0.5\linewidth]{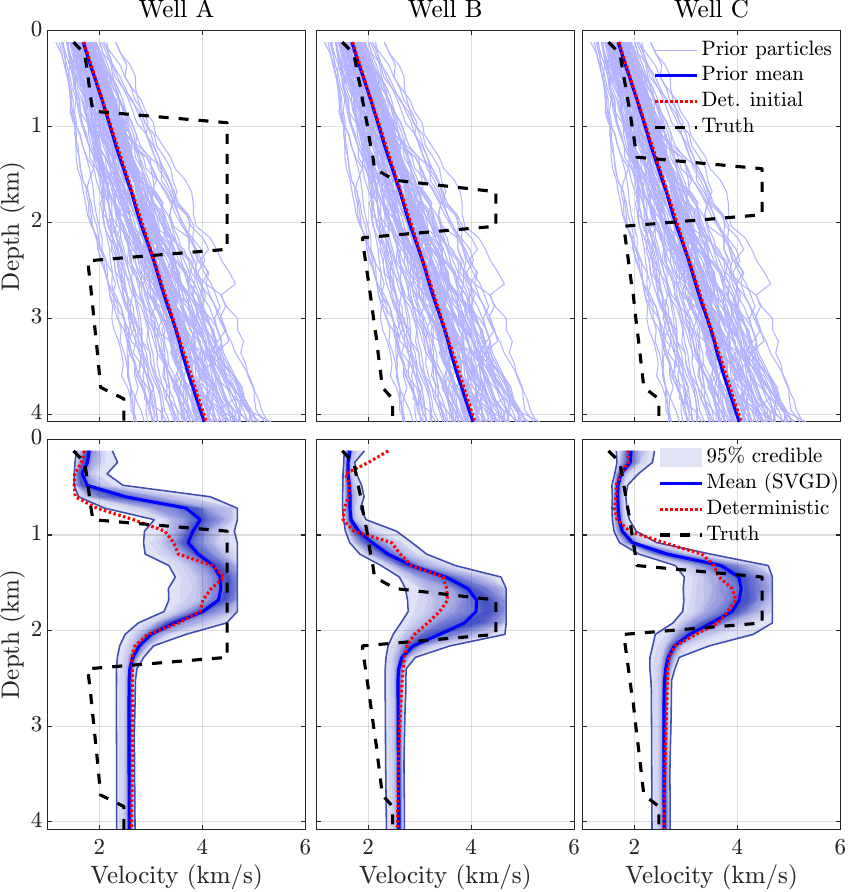}
\caption{SEG/EAGE 3D salt model: well profiles and credible intervals. Vertical velocity profiles at the three well locations (wells A-C) marked in \cref{fig:3D_true_init_det}~(a). Top row: prior particles (thin light-blue curves), their mean (solid blue), the initial model of the deterministic inversion (dotted red), and the true model (dashed black). Bottom row:  95\% credible interval from the final ensemble, posterior mean, and deterministic reconstruction, against the true model (dashed black).  }
    \label{fig:3D_test_CI}
\end{figure}
\begin{figure}
    \centering
    \includegraphics[width=0.9\linewidth]{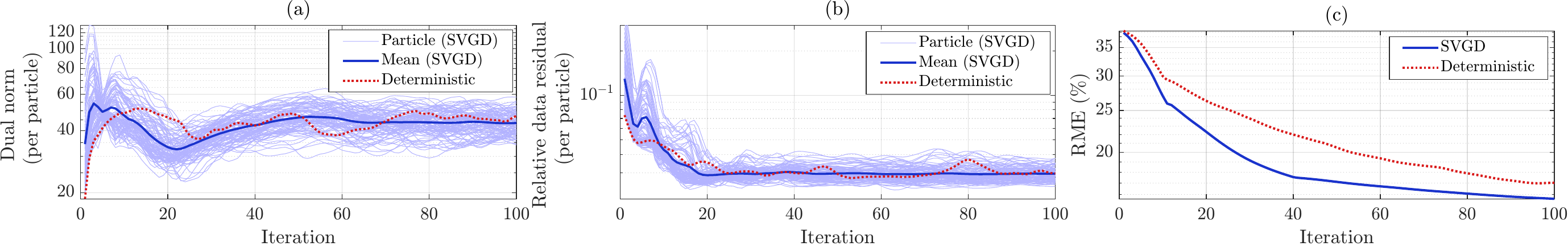}
    \caption{SEG/EAGE 3D salt model: convergence diagnostics. (a-b)  Norm of the data multipliers and relative data residual per particle versus iteration: individual SVGD particles (light blue), ensemble mean (solid blue), and deterministic baseline (dotted red). (c) RME (in percent) of the ensemble mean and of the deterministic reconstruction.}
    \label{fig:3D_erros}
\end{figure}
\subsection{3D problem: the SEG/EAGE 3D salt model}\label{sec:seg3d}
\medskip
We finally apply both algorithms to a 3D problem, the SEG/EAGE salt model \citep{Aminzadeh_1997_DSO}. This example tests whether the data-space worker step~\eqref{eq:smw} and the consensus update remain tractable when the model dimension grows by an order of magnitude relative to the 2D BP example and whether the ensemble still delivers interpretable posterior statistics at a scale where methods such as MCMC are not affordable.  
\subsubsection{Experimental design} \label{sec:seg3d_setup}
\medskip
The true model is a median-filtered version of the SEG/EAGE salt model, covering $13.5 \times 13.5 \times 4$~km, resampled onto a uniform grid of $N_z\times N_x \times N_y=34\times 113 \times 113$ nodes with $120$m spacing, roughly 12 times the size of the BP problem. Velocities range from $1500$~m/s in the water to $4482$~m/s in the salt body, whose irregular crest lies at about $1.0$~km depth (\cref{fig:3D_true_init_det}a).
The acquisition consists of a $10 \times 10$ grid of $N_s=100$ sources ($1.44$~km spacing, $z_s=120$~m) and a fixed $14 \times 11$ grid of $N_r=154$ water-bottom receivers. The location of the sources and receivers is also shown in \cref{fig:3D_true_init_det}a. 
The free parameters in the model subproblem \eqref{m_subprob} are fixed at $\alpha=1$ and $\beta=7$. Additionally, in the SVGD step \eqref{eq:svgd_step} we use $\xi=0.2$ for the first thirty iterations and $\xi=0.05$ thereafter.
The data volume is thus comparable to that of the 2D example but for 12 times the unknowns, which makes this example the more underdetermined of the two by construction. 

The observed data are contaminated with a small amount of Gaussian noise of standard deviation $\sigma_\mathrm{noise}\approx 0.04$~s. The starting model for the deterministic case is the fitted 1D gradient model of \Cref{sec:init}, $v_0(z)=1700+0.6z$~(m/s, $z$ in m), and the $N_p=100$ initial particles are drawn centered on $v_0$. Both algorithms were executed with 100 iterations. For the ADMM-SVGD case, the $N_p\,N_s=10^4$ local subproblems per iteration were distributed over the 30 workers, whereas for the deterministic case there are $N_s=100$ local subproblems.  
%
\subsubsection{Results} \label{sec:seg3d_results}
\medskip
\Cref{fig:3D_true_init_det} compares the true model, the deterministic reconstruction, and the SVGD posterior mean, each with an $X$-$Z$ section at $Y=6.78$~km. Both estimates place the salt body at the correct depth and lateral position and recover the background comparably; the posterior mean is more compact and higher in amplitude at the salt, and beneath it both reconstructions revert to the smooth background trend. Algorithmically, the two runs share the same setup; the main difference is that the SVGD particles are additionally coupled through the RBF kernel in the transport step \eqref{eq:svgd_phi}, whereas the deterministic model follows its own gradient alone. The sharper salt body in the posterior mean may reflect  an ensemble-averaging effect: 100 regularized particles, started from different draws of the same initial distribution and kept apart by the kernel's repulsive term while being pulled toward the data by its attractive term, are averaged at the end of the run, which tends to cancel particle-specific fluctuations that a single deterministic trajectory has no mechanism to cancel. 

The pointwise posterior spread (\cref{fig:3D_svgd}) is concentrated along the salt boundary, which is the deepest interface the acquisition resolves. \Cref{fig:3D_test_CI} shows the same statistics as a function of depth at the three well locations of \cref{fig:3D_true_init_det}a. Down to the top of the salt, the posterior mean tracks the true logs closely, and the 95\% band covers the truth. Beneath the salt, the true velocity drops into a low-velocity zone that neither reconstruction resolves accurately. \\
The relative data residual and the dual-variable norm (\cref{fig:3D_erros}a--b) both drop quickly over the first $\sim\!20$ iterations and then flatten into a plateau for the rest of the run, for all 100 particles and for the deterministic run alike. The RME curve (panel c) is a more informative diagnostic. Both methods reduce it from $\sim 40\%$ to mid-teens over 100 iterations, with the SVGD posterior mean ending below the deterministic estimate.  
\begin{table}
\centering
\caption{Cost of the deterministic and ensemble algorithms for 2D and 3D experiments. Average per-iteration wall-clock time (mean $\pm$ one standard deviation over the 100 outer iterations), total runtime, and peak memory.}
\label{tab:runtime_comparison}
\begin{tabular}{lcccccc}
\toprule
 & \multicolumn{3}{c}{Deterministic ADMM (single model)} & \multicolumn{3}{c}{ADMM-SVGD (100 particles)} \\
\cmidrule(lr){2-4} \cmidrule(lr){5-7}
 & Iter.\ time (s) & Total (s) & Peak mem.\ (GB) & Iter.\ time (s) & Total (s) & Peak mem.\ (GB) \\
\midrule
\multicolumn{7}{l}{Benchmark model} \\
\quad 2D BP (2004) & $0.78 \pm 0.06$ & $79$ & $30.7$ & $50.55 \pm 0.50$ & $5\, 056$ & $40.3$ \\
\quad 3D SEG/EAGE & $6.58 \pm 0.50$ & $660$ & $41.3$ & $449.21 \pm 18.19$ & $44\,922$ & $59.2$ \\
\bottomrule
\end{tabular}
\end{table}
%
%
\subsection{Computational cost and memory}\label{sec:cost}
\medskip
The measured cost of both algorithms on the 2D and 3D benchmarks is summarized in \Cref{tab:runtime_comparison}. 
Two design choices govern the cost of the algorithms. First, the consensus splitting makes the $N_pN_s$ local subproblems of an outer iteration, together with their eikonal solves, mutually independent. Therefore, the ensemble solves a hundred times as many of them as the deterministic implementation: $6700$ against $67$ in 2D and $10^4$ against $100$ in 3D. With both runs distributed over the same pools of 30 workers, the per-iteration wall-clock time grows by a factor of 65 in 2D and 68 in 3D (\Cref{tab:runtime_comparison}), i.e., by less than the hundredfold increase in the number of subproblems. 

Second, the SMW identity \eqref{eq:smw} confines every local solve to a data space of size $N_r^i \times N_r^i$ ($225 \times 225$ in 2D and $154 \times 154$ in 3D), so the source-wise $N \times N$ model-space Hessian is never formed. Only the Jacobians and the small data-space matrices are stored. For 100 particles, the peak memory grows by about 31\% in 2D (from 30.7 to 40.3 GB) and 43\% in 3D (from 41.3 to 59.2 GB). 
%
%
\section{Conclusions}\label{sec_summary}
\medskip
We have presented a distributed proximal SVGD algorithm that makes particle-based Bayesian inference tractable for eikonal-constrained traveltime tomography at the exploration scale. Three ingredients carry the method, and each addresses one of the obstacles identified at the outset. A consensus splitting decomposes the multi-source objective into independent source subproblems and a single regularized master update, in which the non-smooth prior enters through its proximal operator. The Sherman--Morrison--Woodbury identity moves each local Gauss-Newton solve into the data space, replacing the model-space Hessian by a data-space system, whose size is set by the number of receivers. Finally, the particle transport step replaces the deterministic consensus update. Since the local subproblems of different particles are also mutually independent, the particle index becomes a second, embarrassingly parallel dimension rather than a serial multiplier on cost. 

The three benchmarks probe complementary aspects.  On the 1D VSP problem, the ensemble reproduces the depth structure of the pointwise standard deviation of the converged MCMC reference. On the 2D BP 2004 salt model, the posterior mean is largely free of the acquisition footprint left in the deterministic reconstruction and reaches a lower model error. The residual whiteness selection of the penalty parameter brings the data residual to the noise floor without fitting beyond it,  and  the relative model error of the posterior mean grows by less than one percentage point within the noise levels tested. On the 3D SEG/EAGE salt model, the same behavior persists at twelve times the model dimension.  


\section*{Acknowledgements}  
This research was financially supported by the SONATA BIS grant (No. 2022/46/E/ST10/00266) of the National Science Center in Poland.

\section*{Declaration of competing interest}
The authors declare that they have no competing financial interests that could have influenced the work presented in this work. 
\section*{Declaration of generative AI use}
The authors declare that no generative AI tools were used in the preparation of this work. 

\section{Data availability}
The data that support this study are available from the corresponding author on reasonable request. 

\appendix  
\crefalias{section}{appendix} 
\section{Proof of the SMW Matrix Identity}\label{appen_Woodbury}
Let $J\in \mathbb{R}^{N_r\times N}$ with $N_r\ll N$ and $\gamma>0$, we first prove the following identity 
\citep{Guttman_1946_EMC}:
\begin{equation}\label{Woodbury1}
(J^{\top} J + \gamma I_{N})^{-1} J^{\top} = J^{\top} (J J^{\top} + \gamma I_{N_r})^{-1}
\end{equation}
Let us start with a trivial identity $J^{\top} J J^{\top} = J^{\top} J J^{\top}$, then we add $\gamma J^{\top}$ to both side to get
\begin{equation}
    J^{\top} J J^{\top} +\gamma J^{\top}= J^{\top} J J^{\top} +\gamma J^{\top}.
\end{equation}
Factor $J^{\top}$ from the left on the left-hand side and $J^{\top}$ from the right on the right-hand side to obtain 
\begin{equation}\label{eq:I}
    J^{\top} \left(J J^{\top} +\gamma I_{N_r}\right)= \left(J^{\top} J  +\gamma I_{N}\right) J^{\top}.
\end{equation}
Left-multiplying both sides by $\left(J^{\top} J  +\gamma I_{N}\right)^{-1}$ and right-multiplying both sides by $\left(J J^{\top} +\gamma I_{N_r}\right)^{-1}$ yields the identity \eqref{Woodbury1}.

Now we use the identity \eqref{Woodbury1} to prove the identity used in \eqref{eq:smw}:
\begin{equation}\label{Woodbury2}
(J^{\top} J + \gamma I_{N})^{-1} = \frac{1}{\gamma} I_N - \frac{1}{\gamma} J^{\top} (J J^{\top} + \gamma I_{N_r})^{-1} J.
\end{equation}
Starting from the definition of the identity matrix:
\begin{align}
I_{N} & = (J^{\top} J + \gamma I_{N})^{-1} (J^{\top} J + \gamma I_{N})\\
& = (J^{\top} J + \gamma I_{N})^{-1} J^{\top} J + \gamma (J^{\top} J + \gamma I_{N})^{-1},
\end{align}
gives
\begin{align}
\gamma (J^{\top} J + \gamma I_{N})^{-1}
& = I_{N}-(J^{\top} J + \gamma I_{N})^{-1} J^{\top} J\\
& = I_{N}- J^{\top}(JJ^{\top}  + \gamma I_{N_r})^{-1} J,
\end{align}
where the second equality follows from \eqref{eq:I}. Dividing by $\gamma$ gives \eqref{Woodbury2}.
\section{Selection of the penalty parameter by the residual whiteness principle} \label{app:B}



By dividing the objective function in \eqref{local} by $\mu>0$ and defining $\gamma=\alpha / \mu >0$, we get
\begin{equation} \label{local}
\frac{1}{2}\|
 \underbrace{J_i^k p_i-\dobs+d_i^k}_{=:r_i(\gamma)}\|_2^2 + \frac{\gamma}{2} \|p_i - m^k + q_i^k\|_2^2.
\end{equation}
The traveltime residual $r_i(\gamma)$ predicted by the minimizer in $p_i^{k+1}(\gamma)$ in \eqref{pi_k} as a function of $\gamma$ is
\begin{align}
    r_i(\gamma)&=J_i^k  p_i^{k+1}(\gamma) -\dobs +d_i^k \nonumber\\ 
    &=  J_i^k m^k + J_i^k\left((J_i^k)^{\top} J_i^k + \gamma I_N\right)^{-1}\left[(J_i^k)^{\top} (\dobs-G_i(m^k)-\lambda_i^k )- \gamma q_i^k\right] -\dobs +d_i^k \nonumber \\
    &=  J_i^k m^k + \left(J_i^k(J_i^k)^{\top}  + \gamma I_{N_r}\right)^{-1}J_i^k\left[(J_i^k)^{\top} (\dobs-G_i(m^k)-\lambda_i^k )- \gamma q_i^k\right] -\dobs +d_i^k \nonumber\\
    &=  J_i^k m^k + \left(J_i^k(J_i^k)^{\top}  + \gamma I_{N_r}\right)^{-1}\left[J_i^k(J_i^k)^{\top} (\dobs-G_i(m^k)-\lambda_i^k )- \gamma J_i^k q_i^k\right] -\dobs +d_i^k. \label{r1}
\end{align}
where the last equation is obtained using the matrix identity in \eqref{eq:I}.
Adding and subtracting $\gamma \,(\dobs-G_i(m^k)-\lambda_i^k )$ to the term in the braked and using $d_i^k=G_i(m^k)-J_i^k m^k+\lambda_i^k$ from \eqref{AL_p_lin} gives
\begin{align*}
 r_i(\gamma)
   &=-\left(\frac{1}{\gamma}J_i^k(J_i^k)^{\top}  +  I_{N_r}\right)^{-1}\left[\dobs-G_i(m^k)-\lambda_i^k + J_i^k q_i^k\right]
\end{align*}

Following the residual whiteness principle (RWP) \citep{lanza2020residual,pragliola2023admm,aghazade2025automatic}, the penalty parameter is chosen by minimizing the kurtosis of the Fourier transformed residual:
\begin{equation}
    \gamma_i^k =\argmin_{\gamma}~ \frac{ \| \texttt{FFT}[{r_i}(\gamma)] \|_4^4 }{\|\texttt{FFT}[{r_i}(\gamma)] \|_2^4}.
\end{equation}












\bibliographystyle{cas-model2-names}

\bibliography{TT_refs}

\newcommand{\SortNoop}[1]{}
\begin{thebibliography}{63}
\expandafter\ifx\csname natexlab\endcsname\relax\def\natexlab#1{#1}\fi
\providecommand{\url}[1]{\texttt{#1}}
\providecommand{\href}[2]{#2}
\providecommand{\path}[1]{#1}
\providecommand{\DOIprefix}{doi:}
\providecommand{\ArXivprefix}{arXiv:}
\providecommand{\URLprefix}{URL: }
\providecommand{\Pubmedprefix}{pmid:}
\providecommand{\doi}[1]{\href{http://dx.doi.org/#1}{\path{#1}}}
\providecommand{\Pubmed}[1]{\href{pmid:#1}{\path{#1}}}
\providecommand{\bibinfo}[2]{#2}
\ifx\xfnm\relax \def\xfnm[#1]{\unskip,\space#1}\fi
\bibitem[{Aghazade et~al.(2025)Aghazade, Zand and
  Gholami}]{aghazade2025automatic}
\bibinfo{author}{Aghazade, K.}, \bibinfo{author}{Zand, T.},
  \bibinfo{author}{Gholami, A.}, \bibinfo{year}{2025}.
\newblock \bibinfo{title}{Automatic penalty parameter selection by residual
  whiteness principle ({RWP}) and {GCV} for full waveform inversion}.
\newblock \bibinfo{journal}{arXiv preprint arXiv:2512.16757} .
\bibitem[{Almeida and Figueiredo(2013)}]{almeida2013parameter}
\bibinfo{author}{Almeida, M.S.}, \bibinfo{author}{Figueiredo, M.A.},
  \bibinfo{year}{2013}.
\newblock \bibinfo{title}{Parameter estimation for blind and non-blind
  deblurring using residual whiteness measures}.
\newblock \bibinfo{journal}{IEEE Transactions on Image Processing}
  \bibinfo{volume}{22}, \bibinfo{pages}{2751--2763}.
\bibitem[{Aminzadeh et~al.(1997)Aminzadeh, Brac and Kunz}]{Aminzadeh_1997_DSO}
\bibinfo{author}{Aminzadeh, F.}, \bibinfo{author}{Brac, J.},
  \bibinfo{author}{Kunz, T.}, \bibinfo{year}{1997}.
\newblock \bibinfo{title}{{3-D} {Salt} and {Overthrust} models}.
\newblock \bibinfo{publisher}{{SEG/EAGE} 3-{D} {M}odeling {S}eries {N}o.1}.
\bibitem[{Aster et~al.(2018)Aster, Borchers and Thurber}]{Aster_2018_PEI}
\bibinfo{author}{Aster, R.C.}, \bibinfo{author}{Borchers, B.},
  \bibinfo{author}{Thurber, C.H.}, \bibinfo{year}{2018}.
\newblock \bibinfo{title}{Parameter estimation and inverse problems}.
\newblock \bibinfo{publisher}{Elsevier}.
\bibitem[{Billette and Brandsberg-Dahl(2005)}]{billette20052004}
\bibinfo{author}{Billette, F.}, \bibinfo{author}{Brandsberg-Dahl, S.},
  \bibinfo{year}{2005}.
\newblock \bibinfo{title}{The 2004 {BP} velocity benchmark}, in:
  \bibinfo{booktitle}{67th EAGE Conference \& Exhibition},
  \bibinfo{organization}{European Association of Geoscientists \& Engineers}.
  pp. \bibinfo{pages}{cp--1}.
\bibitem[{Bishop et~al.(1985)Bishop, Bube, Cutler, Langan, Love, Resnick,
  Shuey, Spindler and Wyld}]{Bishop_1985_TDV}
\bibinfo{author}{Bishop, T.N.}, \bibinfo{author}{Bube, K.},
  \bibinfo{author}{Cutler, R.T.}, \bibinfo{author}{Langan, R.},
  \bibinfo{author}{Love, P.L.}, \bibinfo{author}{Resnick, J.R.},
  \bibinfo{author}{Shuey, R.}, \bibinfo{author}{Spindler, D.A.},
  \bibinfo{author}{Wyld, H.}, \bibinfo{year}{1985}.
\newblock \bibinfo{title}{Tomographic determination of velocity and depth in
  laterally varying media}.
\newblock \bibinfo{journal}{Geophysics} \bibinfo{volume}{50},
  \bibinfo{pages}{903--923}.
\bibitem[{Boyd et~al.(2010)Boyd, Parikh, Chu, Peleato and
  Eckstein}]{Boyd_2011_DOS}
\bibinfo{author}{Boyd, S.}, \bibinfo{author}{Parikh, N.}, \bibinfo{author}{Chu,
  E.}, \bibinfo{author}{Peleato, B.}, \bibinfo{author}{Eckstein, J.},
  \bibinfo{year}{2010}.
\newblock \bibinfo{title}{Distributed optimization and statistical learning via
  the alternating direction method of multipliers}.
\newblock \bibinfo{journal}{Foundations and trends in machine learning}
  \bibinfo{volume}{3}, \bibinfo{pages}{1--122}.
\bibitem[{B{\"u}rgel et~al.(2017)B{\"u}rgel, Kazimierski and
  Lechleiter}]{Burgel_2017_ASR}
\bibinfo{author}{B{\"u}rgel, F.}, \bibinfo{author}{Kazimierski, K.S.},
  \bibinfo{author}{Lechleiter, A.}, \bibinfo{year}{2017}.
\newblock \bibinfo{title}{A sparsity regularization and total variation based
  computational framework for the inverse medium problem in scattering}.
\newblock \bibinfo{journal}{Journal of Computational Physics}
  \bibinfo{volume}{339}, \bibinfo{pages}{1--30}.
\bibitem[{Combettes and Pesquet(2011)}]{Combettes_2011_PRO}
\bibinfo{author}{Combettes, P.L.}, \bibinfo{author}{Pesquet, J.C.},
  \bibinfo{year}{2011}.
\newblock \bibinfo{title}{Proximal splitting methods in signal processing}, in:
  \bibinfo{editor}{Bauschke, H.H.}, \bibinfo{editor}{Burachik, R.S.},
  \bibinfo{editor}{Combettes, P.L.}, \bibinfo{editor}{Elser, V.},
  \bibinfo{editor}{Luke, D.R.}, \bibinfo{editor}{Wolkowicz, H.} (Eds.),
  \bibinfo{booktitle}{Fixed-Point Algorithms for Inverse Problems in Science
  and Engineering}. \bibinfo{publisher}{Springer New York}.
  volume~\bibinfo{volume}{49} of \textit{\bibinfo{series}{Springer Optimization
  and Its Applications}}, pp. \bibinfo{pages}{185--212}.
\bibitem[{Corrales et~al.(2025)Corrales, Berti, Denel, Williamson, Aleardi and
  Ravasi}]{Corrales_2025_ASV}
\bibinfo{author}{Corrales, M.}, \bibinfo{author}{Berti, S.},
  \bibinfo{author}{Denel, B.}, \bibinfo{author}{Williamson, P.},
  \bibinfo{author}{Aleardi, M.}, \bibinfo{author}{Ravasi, M.},
  \bibinfo{year}{2025}.
\newblock \bibinfo{title}{Annealed {S}tein variational gradient descent for
  improved uncertainty estimation in full-waveform inversion}.
\newblock \bibinfo{journal}{Geophysical Journal International}
  \bibinfo{volume}{241}, \bibinfo{pages}{1088--1113}.
\bibitem[{Curtis and Lomax(2001)}]{Curtis_2001_PIS}
\bibinfo{author}{Curtis, A.}, \bibinfo{author}{Lomax, A.},
  \bibinfo{year}{2001}.
\newblock \bibinfo{title}{Prior information, sampling distributions, and the
  curse of dimensionality}.
\newblock \bibinfo{journal}{Geophysics} \bibinfo{volume}{66},
  \bibinfo{pages}{372--378}.
\bibitem[{Fichtner et~al.(2024)Fichtner, Kennett, Tsai, Thurber, Rodgers, Tape,
  Rawlinson, Borcherdt, Lebedev, Priestley, Morency, Bozdağ, Tromp, Ritsema,
  Romanowicz, Liu, Golos and Lin}]{Fichtner_2024_ST}
\bibinfo{author}{Fichtner, A.}, \bibinfo{author}{Kennett, B.L.N.},
  \bibinfo{author}{Tsai, V.C.}, \bibinfo{author}{Thurber, C.},
  \bibinfo{author}{Rodgers, A.J.}, \bibinfo{author}{Tape, C.},
  \bibinfo{author}{Rawlinson, N.}, \bibinfo{author}{Borcherdt, R.D.},
  \bibinfo{author}{Lebedev, S.}, \bibinfo{author}{Priestley, K.},
  \bibinfo{author}{Morency, C.}, \bibinfo{author}{Bozdağ, E.},
  \bibinfo{author}{Tromp, J.}, \bibinfo{author}{Ritsema, J.},
  \bibinfo{author}{Romanowicz, B.}, \bibinfo{author}{Liu, Q.},
  \bibinfo{author}{Golos, E.}, \bibinfo{author}{Lin, F.}, \bibinfo{year}{2024}.
\newblock \bibinfo{title}{Seismic tomography 2024}.
\newblock \bibinfo{journal}{Bulletin of the Seismological Society of America} .
\bibitem[{Fomel et~al.(2009)Fomel, Luo and Zhao}]{Fomel_2009_FSM}
\bibinfo{author}{Fomel, S.}, \bibinfo{author}{Luo, S.}, \bibinfo{author}{Zhao,
  H.}, \bibinfo{year}{2009}.
\newblock \bibinfo{title}{Fast sweeping method for the factored eikonal
  equation}.
\newblock \bibinfo{journal}{Journal of Computational Physics}
  \bibinfo{volume}{228}, \bibinfo{pages}{6440--6455}.
\bibitem[{Gao and Chen(2025)}]{Gao_2025_LATTE}
\bibinfo{author}{Gao, K.}, \bibinfo{author}{Chen, T.}, \bibinfo{year}{2025}.
\newblock \bibinfo{title}{Latte: Open-source, high-performance traveltime
  computation, tomography, and source location in acoustic and elastic media}.
\newblock \bibinfo{journal}{Geophysical Journal International} .
\bibitem[{Gholami and Aghazade(2024)}]{Gholami_2024_FWI}
\bibinfo{author}{Gholami, A.}, \bibinfo{author}{Aghazade, K.},
  \bibinfo{year}{2024}.
\newblock \bibinfo{title}{Full waveform inversion and lagrange multipliers}.
\newblock \bibinfo{journal}{Geophysical Journal International}
  \bibinfo{volume}{238}, \bibinfo{pages}{109--131}.
\bibitem[{Gholami and Siahkoohi(2010)}]{Gholami_2010_RLN}
\bibinfo{author}{Gholami, A.}, \bibinfo{author}{Siahkoohi, H.},
  \bibinfo{year}{2010}.
\newblock \bibinfo{title}{Regularization of linear and non-linear geophysical
  ill-posed problems with joint sparsity constraints}.
\newblock \bibinfo{journal}{Geophysical Journal International}
  \bibinfo{volume}{180}, \bibinfo{pages}{871--882}.
\bibitem[{Goldstein and Osher(2009)}]{Goldstein_2009_SBM}
\bibinfo{author}{Goldstein, T.}, \bibinfo{author}{Osher, S.},
  \bibinfo{year}{2009}.
\newblock \bibinfo{title}{The split {B}regman method for {L}1-regularized
  problems}.
\newblock \bibinfo{journal}{{SIAM} Journal on Imaging Sciences}
  \bibinfo{volume}{2}, \bibinfo{pages}{323--343}.
\bibitem[{Guttman(1946)}]{Guttman_1946_EMC}
\bibinfo{author}{Guttman, L.}, \bibinfo{year}{1946}.
\newblock \bibinfo{title}{Enlargement methods for computing the inverse
  matrix}.
\newblock \bibinfo{journal}{The Annals of Mathematical Statistics} ,
  \bibinfo{pages}{336--343}.
\bibitem[{Haario et~al.(2001)Haario, Saksman and Tamminen}]{Haario_2001_AAM}
\bibinfo{author}{Haario, H.}, \bibinfo{author}{Saksman, E.},
  \bibinfo{author}{Tamminen, J.}, \bibinfo{year}{2001}.
\newblock \bibinfo{title}{An adaptive {M}etropolis algorithm} .
\bibitem[{Jin and Zou(2010)}]{Jin_2010_HBI}
\bibinfo{author}{Jin, B.}, \bibinfo{author}{Zou, J.}, \bibinfo{year}{2010}.
\newblock \bibinfo{title}{Hierarchical {B}ayesian inference for ill-posed
  problems via variational method}.
\newblock \bibinfo{journal}{Journal of Computational Physics}
  \bibinfo{volume}{229}, \bibinfo{pages}{7317--7343}.
\bibitem[{Kira and Noir(2025)}]{Kira_2025_FWA}
\bibinfo{author}{Kira, L.}, \bibinfo{author}{Noir, J.}, \bibinfo{year}{2025}.
\newblock \bibinfo{title}{Full-waveform acoustic tomography for fluid
  temperature and flow}.
\newblock \bibinfo{journal}{Experiments in Fluids} \bibinfo{volume}{66},
  \bibinfo{pages}{145}.
\bibitem[{Klibanov et~al.(2023)Klibanov, Li and Zhang}]{KLIBANOV_2023_NS3D}
\bibinfo{author}{Klibanov, M.V.}, \bibinfo{author}{Li, J.},
  \bibinfo{author}{Zhang, W.}, \bibinfo{year}{2023}.
\newblock \bibinfo{title}{Numerical solution of the 3-{D} travel time
  tomography problem}.
\newblock \bibinfo{journal}{Journal of Computational Physics}
  \bibinfo{volume}{476}, \bibinfo{pages}{111910}.
\bibitem[{Lanza et~al.(2020)Lanza, Pragliola, Sgallari
  et~al.}]{lanza2020residual}
\bibinfo{author}{Lanza, A.}, \bibinfo{author}{Pragliola, M.},
  \bibinfo{author}{Sgallari, F.}, et~al., \bibinfo{year}{2020}.
\newblock \bibinfo{title}{Residual whiteness principle for parameter-free image
  restoration}.
\newblock \bibinfo{journal}{Electronic Transactions on Numerical Analysis}
  \bibinfo{volume}{53}, \bibinfo{pages}{329--351}.
\bibitem[{Li et~al.(2025)Li, Zhang, Zhu and Gao}]{Li_2025_STT}
\bibinfo{author}{Li, Y.}, \bibinfo{author}{Zhang, Y.}, \bibinfo{author}{Zhu,
  X.}, \bibinfo{author}{Gao, J.}, \bibinfo{year}{2025}.
\newblock \bibinfo{title}{Seismic traveltime tomography based on ensemble
  {K}alman inversion}.
\newblock \bibinfo{journal}{Geophysical Journal International}
  \bibinfo{volume}{240}, \bibinfo{pages}{290--302}.
\bibitem[{Liu and Wang(2016)}]{Liu_2016_SVGD}
\bibinfo{author}{Liu, Q.}, \bibinfo{author}{Wang, D.}, \bibinfo{year}{2016}.
\newblock \bibinfo{title}{Stein variational gradient descent: A general purpose
  {B}ayesian inference algorithm}, in: \bibinfo{booktitle}{Advances in Neural
  Information Processing Systems}.
\bibitem[{Loris et~al.(2010)Loris, Douma, Nolet, Daubechies and
  Regone}]{Loris_2010_NRT}
\bibinfo{author}{Loris, I.}, \bibinfo{author}{Douma, H.},
  \bibinfo{author}{Nolet, G.}, \bibinfo{author}{Daubechies, I.},
  \bibinfo{author}{Regone, C.}, \bibinfo{year}{2010}.
\newblock \bibinfo{title}{Nonlinear regularization techniques for seismic
  tomography}.
\newblock \bibinfo{journal}{Journal of Computational Physics}
  \bibinfo{volume}{229}, \bibinfo{pages}{890--905}.
\bibitem[{Loris et~al.(2007)Loris, Nolet, {Daubechies} and
  Dahlen}]{Loris_2007_TIU}
\bibinfo{author}{Loris, I.}, \bibinfo{author}{Nolet, G.},
  \bibinfo{author}{{Daubechies}, I.}, \bibinfo{author}{Dahlen, F.A.},
  \bibinfo{year}{2007}.
\newblock \bibinfo{title}{Tomographic inversion using $\ell_1$-norm
  regularization of wavelet coefficients}.
\newblock \bibinfo{journal}{Geophysical Journal International}
  \bibinfo{volume}{170}, \bibinfo{pages}{359--370}.
\bibitem[{Martin et~al.(2012)Martin, Wilcox, Burstedde and
  Ghattas}]{Martin_2012_SNM}
\bibinfo{author}{Martin, J.}, \bibinfo{author}{Wilcox, L.},
  \bibinfo{author}{Burstedde, C.}, \bibinfo{author}{Ghattas, O.},
  \bibinfo{year}{2012}.
\newblock \bibinfo{title}{A stochastic {N}ewton {MCMC} method for large-scale
  statistical inverse problems with application to seismic inversion}.
\newblock \bibinfo{journal}{{SIAM} Journal of Scientific Computing}
  \bibinfo{volume}{34(3)}, \bibinfo{pages}{A1460--A1487}.
\bibitem[{Miele et~al.(1972)Miele, Moseley, Levy and Coggins}]{Miele_1972_OTM}
\bibinfo{author}{Miele, A.}, \bibinfo{author}{Moseley, P.},
  \bibinfo{author}{Levy, A.}, \bibinfo{author}{Coggins, G.},
  \bibinfo{year}{1972}.
\newblock \bibinfo{title}{On the method of multipliers for mathematical
  programming problems}.
\newblock \bibinfo{journal}{Journal of optimization Theory and Applications}
  \bibinfo{volume}{10}, \bibinfo{pages}{1--33}.
\bibitem[{Mosegaard and Tarantola(1995)}]{Mosegaard_1995_MCS}
\bibinfo{author}{Mosegaard, K.}, \bibinfo{author}{Tarantola, A.},
  \bibinfo{year}{1995}.
\newblock \bibinfo{title}{{M}onte {C}arlo sampling of solutions to inverse
  problems}.
\newblock \bibinfo{journal}{Journal of Geophysical Research}
  \bibinfo{volume}{100 B7}, \bibinfo{pages}{12431--12447}.
\bibitem[{Nocedal and Wright(2006)}]{Nocedal_2006_NO}
\bibinfo{author}{Nocedal, J.}, \bibinfo{author}{Wright, S.J.},
  \bibinfo{year}{2006}.
\newblock \bibinfo{title}{Numerical Optimization}.
\newblock \bibinfo{edition}{2\textsuperscript{nd}} ed.,
  \bibinfo{publisher}{Springer}.
\bibitem[{Nolet(2008)}]{Nolet_2008_BST}
\bibinfo{author}{Nolet, G.}, \bibinfo{year}{2008}.
\newblock \bibinfo{title}{A Breviary of Seismic Tomography}.
\newblock \bibinfo{publisher}{Cambridge University Press},
  \bibinfo{address}{Cambridge, UK}.
\bibitem[{Operto et~al.(2023)Operto, Gholami, Aghamiry, Guo, Beller, Aghazade,
  Mamfoumbi, Combe and Ribodetti}]{Operto_2023_ESS}
\bibinfo{author}{Operto, S.}, \bibinfo{author}{Gholami, A.},
  \bibinfo{author}{Aghamiry, H.S.}, \bibinfo{author}{Guo, G.},
  \bibinfo{author}{Beller, S.}, \bibinfo{author}{Aghazade, K.},
  \bibinfo{author}{Mamfoumbi, F.}, \bibinfo{author}{Combe, L.},
  \bibinfo{author}{Ribodetti, A.}, \bibinfo{year}{2023}.
\newblock \bibinfo{title}{Extending the search space of full-waveform inversion
  beyond the single-scattering {B}orn approximation: A tutorial review}.
\newblock \bibinfo{journal}{Geophysics} \bibinfo{volume}{88},
  \bibinfo{pages}{1--32}.
\bibitem[{Othmani et~al.(2023)Othmani, Dokhanchi, Merchel, Vogel, Altinsoy,
  Voelker and Takali}]{Othmani_2023_ATR}
\bibinfo{author}{Othmani, C.}, \bibinfo{author}{Dokhanchi, N.S.},
  \bibinfo{author}{Merchel, S.}, \bibinfo{author}{Vogel, A.},
  \bibinfo{author}{Altinsoy, M.E.}, \bibinfo{author}{Voelker, C.},
  \bibinfo{author}{Takali, F.}, \bibinfo{year}{2023}.
\newblock \bibinfo{title}{Acoustic tomographic reconstruction of temperature
  and flow fields with focus on atmosphere and enclosed spaces: A review}.
\newblock \bibinfo{journal}{Applied Thermal Engineering} \bibinfo{volume}{223},
  \bibinfo{pages}{119953}.
\bibitem[{Powell(1969)}]{Powell_1969_NLC}
\bibinfo{author}{Powell, M.J.}, \bibinfo{year}{1969}.
\newblock \bibinfo{title}{A method for nonlinear constraints in minimization
  problems}.
\newblock \bibinfo{journal}{Optimization} , \bibinfo{pages}{283--298}.
\bibitem[{Pragliola et~al.(2023)Pragliola, Calatroni, Lanza and
  Sgallari}]{pragliola2023admm}
\bibinfo{author}{Pragliola, M.}, \bibinfo{author}{Calatroni, L.},
  \bibinfo{author}{Lanza, A.}, \bibinfo{author}{Sgallari, F.},
  \bibinfo{year}{2023}.
\newblock \bibinfo{title}{{ADMM}-based residual whiteness principle for
  automatic parameter selection in single image super-resolution problems}.
\newblock \bibinfo{journal}{Journal of Mathematical Imaging and Vision}
  \bibinfo{volume}{65}, \bibinfo{pages}{99--123}.
\bibitem[{Raanes(2011)}]{Patrick_2011_FM}
\bibinfo{author}{Raanes, P.N.}, \bibinfo{year}{2011}.
\newblock \bibinfo{title}{patricknraanes/fm: Version 1.0}.
\newblock \URLprefix \url{https://doi.org/10.5281/zenodo.2025811},
  \DOIprefix\doi{10.5281/zenodo.2025811}.
\bibitem[{Rawlinson and Spakman(2016)}]{Rawlinson_2016_OUS}
\bibinfo{author}{Rawlinson, N.}, \bibinfo{author}{Spakman, W.},
  \bibinfo{year}{2016}.
\newblock \bibinfo{title}{On the use of sensitivity tests in seismic
  tomography}.
\newblock \bibinfo{journal}{Geophysical Journal International}
  \bibinfo{volume}{205}, \bibinfo{pages}{1221--1243}.
\bibitem[{Rizzuti et~al.(2020)Rizzuti, Siahkoohi, Witte and
  Herrmann}]{Rizzuti_2020_PUD}
\bibinfo{author}{Rizzuti, G.}, \bibinfo{author}{Siahkoohi, A.},
  \bibinfo{author}{Witte, P.A.}, \bibinfo{author}{Herrmann, F.J.},
  \bibinfo{year}{2020}.
\newblock \bibinfo{title}{Parameterizing uncertainty by deep invertible
  networks, an application to reservoir characterization}, in:
  \bibinfo{booktitle}{Society of Exploration Geophysicists Technical Program
  Expanded Abstracts}, pp. \bibinfo{pages}{1541--1545}.
\bibitem[{Rudin et~al.(1992)Rudin, Osher and Fatemi}]{Rudin_1992_NTV}
\bibinfo{author}{Rudin, L.}, \bibinfo{author}{Osher, S.},
  \bibinfo{author}{Fatemi, E.}, \bibinfo{year}{1992}.
\newblock \bibinfo{title}{Nonlinear total variation based noise removal
  algorithms}.
\newblock \bibinfo{journal}{Physica D} \bibinfo{volume}{60},
  \bibinfo{pages}{259--268}.
\bibitem[{Ryberg and Haberland(2018)}]{Ryberg_2018_BIR}
\bibinfo{author}{Ryberg, T.}, \bibinfo{author}{Haberland, C.},
  \bibinfo{year}{2018}.
\newblock \bibinfo{title}{Bayesian inversion of refraction seismic traveltime
  data}.
\newblock \bibinfo{journal}{Geophysical Journal International}
  \bibinfo{volume}{214}, \bibinfo{pages}{1626--1642}.
\bibitem[{Sethian(1999)}]{Sethian_1999_FMM}
\bibinfo{author}{Sethian, J.A.}, \bibinfo{year}{1999}.
\newblock \bibinfo{title}{Fast marching methods}.
\newblock \bibinfo{journal}{SIAM review} \bibinfo{volume}{41},
  \bibinfo{pages}{199--235}.
\bibitem[{Shin and Shutin(2022)}]{Shin_2022_DTT}
\bibinfo{author}{Shin, B.S.}, \bibinfo{author}{Shutin, D.},
  \bibinfo{year}{2022}.
\newblock \bibinfo{title}{Distributed traveltime tomography using kernel-based
  regression in seismic networks}.
\newblock \bibinfo{journal}{IEEE Geoscience and Remote Sensing Letters}
  \bibinfo{volume}{19}, \bibinfo{pages}{1--5}.
\bibitem[{Si et~al.(2025)Si, Philip, Wei and Qian}]{Si_2025_HoAA}
\bibinfo{author}{Si, B.}, \bibinfo{author}{Philip, A.}, \bibinfo{author}{Wei,
  Y.}, \bibinfo{author}{Qian, J.}, \bibinfo{year}{2025}.
\newblock \bibinfo{title}{High-order accurate adjoint-state methods for
  three-dimensional high-resolution first-arrival traveltime tomography}.
\newblock \bibinfo{journal}{Journal of Computational Physics}
  \bibinfo{volume}{524}, \bibinfo{pages}{113715}.
\bibitem[{Siahkoohi et~al.(2026)Siahkoohi, Aghazade and
  Gholami}]{Siahkoohi_2026_DSP}
\bibinfo{author}{Siahkoohi, A.}, \bibinfo{author}{Aghazade, K.},
  \bibinfo{author}{Gholami, A.}, \bibinfo{year}{2026}.
\newblock \bibinfo{title}{Dual-space posterior sampling for bayesian inference
  in constrained inverse problems}.
\newblock \bibinfo{journal}{arXiv preprint arXiv:2603.00393} .
\bibitem[{Slotnick and Geyer(1959)}]{Slotnick_1959_ETaP}
\bibinfo{author}{Slotnick, M.M.}, \bibinfo{author}{Geyer, R.A.},
  \bibinfo{year}{1959}.
\newblock \bibinfo{title}{Lessons in Seismic Computing}.
\newblock \bibinfo{publisher}{Society of Exploration Geophysicists}.
\bibitem[{Tarantola(2005)}]{Tarantola_2005_IPT}
\bibinfo{author}{Tarantola, A.}, \bibinfo{year}{2005}.
\newblock \bibinfo{title}{Inverse Problem Theory and Methods for Model
  Parameter Estimation}.
\newblock \bibinfo{publisher}{Society for {I}ndustrial and {A}pplied
  {M}athematics}, \bibinfo{address}{Philadelphia}.
\bibitem[{Tikhonov and Arsenin(1977)}]{Tikhonov_1977_SIP}
\bibinfo{author}{Tikhonov, A.}, \bibinfo{author}{Arsenin, V.},
  \bibinfo{year}{1977}.
\newblock \bibinfo{title}{Solution of ill-posed problems}.
\newblock \bibinfo{publisher}{Winston, Washington, {DC}}.
\bibitem[{Treister and Haber(2016)}]{Treister_2016_FMA}
\bibinfo{author}{Treister, E.}, \bibinfo{author}{Haber, E.},
  \bibinfo{year}{2016}.
\newblock \bibinfo{title}{A fast marching algorithm for the factored eikonal
  equation}.
\newblock \bibinfo{journal}{Journal of Computational physics}
  \bibinfo{volume}{324}, \bibinfo{pages}{210--225}.
\bibitem[{Tsianos et~al.(2012)Tsianos, Lawlor and Rabbat}]{Tsianos_2012_CBD}
\bibinfo{author}{Tsianos, K.I.}, \bibinfo{author}{Lawlor, S.},
  \bibinfo{author}{Rabbat, M.G.}, \bibinfo{year}{2012}.
\newblock \bibinfo{title}{Consensus-based distributed optimization: Practical
  issues and applications in large-scale machine learning},
  \bibinfo{organization}{IEEE}. pp. \bibinfo{pages}{1543--1550}.
\bibitem[{Vehtari et~al.(2021)Vehtari, Gelman, Simpson, Carpenter and
  B{\"u}rkner}]{Vehtari_2021_RNF}
\bibinfo{author}{Vehtari, A.}, \bibinfo{author}{Gelman, A.},
  \bibinfo{author}{Simpson, D.}, \bibinfo{author}{Carpenter, B.},
  \bibinfo{author}{B{\"u}rkner, P.C.}, \bibinfo{year}{2021}.
\newblock \bibinfo{title}{Rank-normalization, folding, and localization: {A}n
  improved $\hat{R}$ for assessing convergence of {MCMC} (with discussion)}.
\newblock \bibinfo{journal}{Bayesian analysis} \bibinfo{volume}{16},
  \bibinfo{pages}{667--718}.
\bibitem[{Virieux and Operto(2009)}]{Virieux_2009_OFW}
\bibinfo{author}{Virieux, J.}, \bibinfo{author}{Operto, S.},
  \bibinfo{year}{2009}.
\newblock \bibinfo{title}{An overview of full waveform inversion in exploration
  geophysics}.
\newblock \bibinfo{journal}{Geophysics} \bibinfo{volume}{74},
  \bibinfo{pages}{WCC1--WCC26}.
\bibitem[{Wahba(1990)}]{wahba1990spline}
\bibinfo{author}{Wahba, G.}, \bibinfo{year}{1990}.
\newblock \bibinfo{title}{Spline models for observational data}.
\newblock \bibinfo{publisher}{SIAM}.
\bibitem[{Yin et~al.(2025)Yin, Orozco, Louboutin and Herrmann}]{Yin_2025_MVI}
\bibinfo{author}{Yin, Z.}, \bibinfo{author}{Orozco, R.},
  \bibinfo{author}{Louboutin, M.}, \bibinfo{author}{Herrmann, F.J.},
  \bibinfo{year}{2025}.
\newblock \bibinfo{title}{{WISER}: {M}ultimodal variational inference for
  full-waveform inversion without dimensionality reduction}.
\newblock \bibinfo{journal}{Geophysics} \bibinfo{volume}{90},
  \bibinfo{pages}{A1--A7}.
\bibitem[{Zand et~al.(2020)Zand, Siahkoohi, Malcolm, Gholami and
  Richardson}]{Zand_2020_COT}
\bibinfo{author}{Zand, T.}, \bibinfo{author}{Siahkoohi, H.R.},
  \bibinfo{author}{Malcolm, A.}, \bibinfo{author}{Gholami, A.},
  \bibinfo{author}{Richardson, A.}, \bibinfo{year}{2020}.
\newblock \bibinfo{title}{Consensus optimization of total variation--based
  reverse time migration}.
\newblock \bibinfo{journal}{Computational Geosciences} \bibinfo{volume}{24},
  \bibinfo{pages}{1393--1407}.
\bibitem[{Zhang et~al.(1998)Zhang, {ten Brink} and {Toks\"
  oz}}]{Zhang_1998_NRR}
\bibinfo{author}{Zhang, J.}, \bibinfo{author}{{ten Brink}, U.S.},
  \bibinfo{author}{{Toks\" oz}, M.N.}, \bibinfo{year}{1998}.
\newblock \bibinfo{title}{Nonlinear refraction and reflection travel time
  tomography}.
\newblock \bibinfo{journal}{Journal of Geophysical Research}
  \bibinfo{volume}{103}, \bibinfo{pages}{29,743--29,757}.
\bibitem[{Zhang and Toks\"{o}z(1998)}]{Zhang_1998_NRT}
\bibinfo{author}{Zhang, J.}, \bibinfo{author}{Toks\"{o}z, M.N.},
  \bibinfo{year}{1998}.
\newblock \bibinfo{title}{Nonlinear refraction traveltime tomography}.
\newblock \bibinfo{journal}{Geophysics} \bibinfo{volume}{63},
  \bibinfo{pages}{1726--1737}.
\bibitem[{Zhang and Curtis(2020)}]{Zhang_2020_VFW}
\bibinfo{author}{Zhang, X.}, \bibinfo{author}{Curtis, A.},
  \bibinfo{year}{2020}.
\newblock \bibinfo{title}{Variational full-waveform inversion}.
\newblock \bibinfo{journal}{Geophysical Journal International}
  \bibinfo{volume}{222}, \bibinfo{pages}{406--411}.
\bibitem[{Zhang et~al.(2023)Zhang, Lomas, Zhou, Zheng and
  Curtis}]{Zhang_2023_3DB}
\bibinfo{author}{Zhang, X.}, \bibinfo{author}{Lomas, A.},
  \bibinfo{author}{Zhou, M.}, \bibinfo{author}{Zheng, Y.},
  \bibinfo{author}{Curtis, A.}, \bibinfo{year}{2023}.
\newblock \bibinfo{title}{3-{D} {B}ayesian variational full waveform
  inversion}.
\newblock \bibinfo{journal}{Geophysical Journal International}
  \bibinfo{volume}{234}, \bibinfo{pages}{546--561}.
\bibitem[{Zhao(2005)}]{Zhao_2005_FSM}
\bibinfo{author}{Zhao, H.}, \bibinfo{year}{2005}.
\newblock \bibinfo{title}{A fast sweeping method for eikonal equations}.
\newblock \bibinfo{journal}{Mathematics of computation} \bibinfo{volume}{74},
  \bibinfo{pages}{603--627}.
\bibitem[{Zhao and Curtis(2024)}]{Zhao_2024_PSV}
\bibinfo{author}{Zhao, X.}, \bibinfo{author}{Curtis, A.}, \bibinfo{year}{2024}.
\newblock \bibinfo{title}{Physically structured variational inference for
  {B}ayesian full waveform inversion}.
\newblock \bibinfo{journal}{Journal of Geophysical Research: Solid Earth}
  \bibinfo{volume}{129}.
\bibitem[{Zhao and Curtis(2025)}]{Zhao_2025_EBF}
\bibinfo{author}{Zhao, X.}, \bibinfo{author}{Curtis, A.}, \bibinfo{year}{2025}.
\newblock \bibinfo{title}{Efficient {B}ayesian full-waveform inversion and
  analysis of prior hypotheses in three dimensions}.
\newblock \bibinfo{journal}{Geophysics} \bibinfo{volume}{90},
  \bibinfo{pages}{R373--R388}.
\bibitem[{Zhao et~al.(2022)Zhao, Curtis and Zhang}]{Zhao_2022_BST}
\bibinfo{author}{Zhao, X.}, \bibinfo{author}{Curtis, A.},
  \bibinfo{author}{Zhang, X.}, \bibinfo{year}{2022}.
\newblock \bibinfo{title}{Bayesian seismic tomography using normalizing flows}.
\newblock \bibinfo{journal}{Geophysical Journal International}
  \bibinfo{volume}{228}, \bibinfo{pages}{213--239}.

\end{thebibliography}



\end{document}